\documentclass[12pt,preprint]{aastex}

\shorttitle{Study of time lags in HETE-2 Gamma-Ray Bursts}
\shortauthors{J. Bolmont et al.}

\begin{document}

\title{Study of time lags in HETE-2 Gamma-Ray Bursts with redshift:\\
search for astrophysical effects and Quantum Gravity signature}

\author{
J. Bolmont$^{1,4,\star}$,  
A. Jacholkowska$^{1}$, 
J.-L. Atteia$^{2}$,
F. Piron$^{1}$ and 
G. Pizzichini$^{3}$
}

\altaffiltext{1}{LPTA, Universit\'e Montpellier 2, CNRS/IN2P3, Montpellier, France}
\altaffiltext{2}{LAOMP, Universit\'e Paul Sabatier, CNRS/INSU, Toulouse, France}
\altaffiltext{3}{INAF/IASF, Bologna, Italy}
\altaffiltext{4}{present address: DESY, D-15738 Zeuthen, Germany}
\altaffiltext{$\star$}{\texttt{Julien.Bolmont@desy.de}}

\begin{abstract}
The study of time lags between spikes in Gamma-Ray Burst light curves in different energy bands as a function of redshift may lead to the detection of effects due to Quantum Gravity. We present an analysis of 15 Gamma-Ray Bursts with measured redshift, detected by the HETE-2 mission in order to measure time lags related to astrophysical effects and search for Quantum Gravity signature in the framework of an extra-dimension string model.

The wavelet transform method is used both for de-noising the light curves and for the detection of sharp transitions. The use of photon tagged data allows us to consider various energy ranges and to evaluate systematic effects due to selections and cuts.

The analysis of maxima and minima of the light curves leads to no significant Quantum Gravity effect. A lower limit at 95\% Confidence Level on the Quantum Gravity scale parameter of $2\times10^{15}$~GeV is set.
\end{abstract}



\section{Introduction}\label{sec:intro}

Particle Physics provides a challenging description of Quantum Gravity
in the framework of the String Theory \citep{ellis1,ellis2,amelino}. Within this scheme, gravitation is considered as a gauge interaction and Quantum Gravity effects result from graviton-like exchange in a background classical space-time.
This approach does not imply a ``spontaneous'' Lorentz symmetry breaking,
as it may appear in General Relativity with Loop Quantum Gravity, which postulates discrete space-time in the Planckian regime \citep{gambini,smolin,alfaro,alfaro2}. 

In recent years, several experimental probes have been proposed to test Lorentz invariance in the frame of both particle physics and astrophysics (see \citet{mattingly} and \citet{sarkar} for a review). In the domain of gamma-ray astronomy, the idea proposed by \citet{amelino1} to use Gamma-Ray Bursts (GRBs) to measure arrival time of photons of different energies was taken further and applied to other sources like pulsars or blazars. \citet{kaaret} gets a limit on the quantum gravity energy scale of $1.8\times10^{15}$~GeV using the Crab pulsar observed by EGRET. \citet{biller} use a gamma-ray flare from Markarian 421 and obtain a limit of $6\times10^{16}$~GeV.

GRBs are the most distant variable sources detected by present experiments in the energy range from keV to~GeV. These violent and explosive events are followed by a delayed emission (an afterglow) at radio, infrared, visible and X-ray wavelengths. The energy released during the explosion phase is of the order of $10^{51}$ erg when the beaming corrections are applied.

As mentioned above, GRBs have been proposed to study modification of photon propagation implied by quantum gravity. They are good candidates for this kind of work since they are bright transient sources located at cosmological distances. Some studies use only one GRB to get a limit on the quantum energy scale. \citet{boggs} obtain a limit of $1.8\times10^{17}$~GeV using GRB 021206 observed by RHESSI. This is currently the best limit obtained with a GRB but it relies on an estimation of the redshift which suffers from a big uncertainty. With GRB 930131, \citet{schaefer99} gets a limit of $8.3\times10^{16}$~GeV with 30 keV to 80~MeV photons. Finally, \citet{ellis,ellis5} make use of several GRBs at different distances with measured redshifts. Using 9 GRBs observed by BATSE and OSSE and 35~GRBs seen by BATSE, HETE-2 and SWIFT, they obtain limits which are respectively $6\times10^{15}$~GeV and $9\times10^{15}$~GeV. \citet{ellis5} use public data of HETE, which is available in 3 fixed energy bands. In the present paper, we use photon tagged data, which allow us to study several energy gap scenarios. All papers quoted in this paragraph make use of the model described in \citep{ellis3}. However, a different formalism has been proposed by \citet{rodriguez} which leads to a limit of $6.6\times10^{16}$~GeV, after analysis of GRB~051221A.

While GRBs constitute interesting sources to test the Lorentz violation, they are not perfect signals. The GRB prompt emission extends over many decades in energy (from the optical to~GeV) and it is conceivable that the emission at very different wavelengths (e.g. optical and gamma-rays) is produced by different mechanisms, resulting in different light curves.
Until we fully understand the radiative transfer in GRBs, we must restrict ourselves to an energy domain where the emission is produced by a single process. This is the case for the prompt X-ray and low energy gamma-ray emissions which have very similar light curves and are thus appropriate for the study of Lorentz violation. The influence of the source effects in the present analysis will be addressed later in this article.

In order to search for Quantum Gravity using the GRBs detected by the HETE-2 mission, we concentrate on the extra-dimension string model interpretation. The employed model \citep{ellis3} relies on the assumption that photons propagate in the vacuum which may exhibit a non-trivial refractive index due to its foamy structure on a characteristic scale approaching the Planck length $l \sim m^{-1}_\mathrm{Planck}$. This would imply light velocity variation as a function of the energy of the photon ($E$). In particular, the effects of Quantum Gravity on the light group velocity $v$ would lead to:

\begin{equation}
\label{form:ve}
v(E) = \frac{c}{n(E)},
\end{equation}
where $n(E)$ is the refractive index of the foam. Generally, the Quantum Gravity energy scale $\mathrm{E}_\mathrm{QG}$ is considered to be close to the Planck scale. This allows us to represent the standard photon dispersion relation with $\mathrm{E}/\mathrm{E}_\mathrm{QG}$ expansion:

\begin{equation}
\label{form:di}
c^2p^2 = E^2 \left(1 + \xi \frac{E}{\mathrm{E}_\mathrm{QG}} + O(\frac{E^2}{\mathrm{E}_\mathrm{QG}^2})\right)\ \ \mathrm{and}\ \ v(E) \approx c\left(1-\xi\frac{E}{\mathrm{E}_\mathrm{QG}}\right),
\end{equation}
where $\xi$ is a model parameter whose value is set to 1 in the following \citep{amelino1}.

The analysis of time lags as a function of redshift requires a correction due to the expansion of the Universe, which depends on the cosmological model. Following the analysis of the BATSE data and more recently of the HETE-2 and SWIFT GRB data \citep{ellis,ellis5} and considering the Standard Cosmological Model \citep{bahcall} with flat expanding Universe and a cosmological constant, the difference in the arrival time of two photons with energy difference $\Delta E$ is given by the formula:

\begin{equation}
\label{form:dtc}
\Delta t = \mathrm{H}_0^{-1} {\frac{\Delta E}{\mathrm{E}_\mathrm{QG}}} \int_{0}^{z} \frac{(1 + z)\,dz}{h(z)},
\end{equation}
 
\noindent
where
\begin{equation}
\label{form:hz}
h(z) = \sqrt{\Omega_\Lambda + \Omega_\mathrm{M} (1+z)^3}.
\end{equation}
\noindent
We assume $\Omega_{tot} = \Omega_\Lambda + \Omega_\mathrm{M} = 1$, $\Omega_\Lambda = 0.7$ and $\mathrm{H}_{0} = 71\:\mathrm{km}\:\mathrm{s}^{-1}\:\mathrm{Mpc}^{-1}$.

Equation~\ref{form:dtc} is different from the one used in \citet{ellis,ellis5} where the authors computed the time lag at redshift $z$. As the time lag is measured at redshift 0, a factor $(1 + z)$ has been introduced in the integral. A more detailed computation leading to Equation~\ref{form:dtc} is given in Appendix \ref{an}.

In order to probe the energy dependence of the velocity of light induced by Quantum Gravity, we analyse the time lags as a function of the redshift.
Possible effects intrinsic to the astrophysical sources could also produce energy dependent time lags. The analysis as a function of $z$ ensures, in principle, that the results are independent of such effects. At the first order of the dispersion relation, we fit the evolution of the time lags as a function of $z$:

\begin{equation}
\label{form:linformab1}
< \Delta t >\ =\ \mathrm{a}\,K_l(z) + \mathrm{b}\,(1+z),
\end{equation}
where 
\begin{equation}
\label{form:kl}
K_l(z) = \int_{0}^{z} \frac{(1 + z)\,dz}{h(z)}.
\end{equation}
and where parameters $a$ and $b$  stand for extrinsic (Quantum Gravity) and intrinsic effects respectively. Provided the change in the definition of $K_l$, this formulation is the same as the one used by \citet{ellis5}. It differs from \citet{ellis} with respect to $b$ parameter which is here expressed in the source frame of reference instead of the observer frame.

In this paper, following \citet{ellis,ellis5}, we will apply the wavelet transform methods for noise removal and for high accuracy timings of sharp transients in 15 GRB light curves measured by the on-board FREGATE detector of the HETE-2 mission, assorted with redshift values given by the optical observations of their afterglows. The study of time lags between photons in various energy bands will allow us to constrain the Quantum Gravity scale in the linear string model as discussed above. The astrophysical effects detected in previous studies are also considered briefly.

The layout of the paper is as follows: after a brief description of the HETE-2 experiment and gamma measurements in Section~\ref{sec:hete2}, we present in Section~\ref{sec:method} the methods for de-noising and for the search for sharp transitions in the light curves. The results on Quantum Gravity scale determination are given in Section~\ref{sec:res} and possible effects other than those produced by the Quantum Gravity (systematic effects in the proposed analysis and astrophysical source effects) are presented in Section~\ref{sec:syseffect}. Finally, the overall discussion of the results and their possible interpretations is presented in Section~\ref{sec:summary}.

\section{HETE-2 Experiment}\label{sec:hete2}

The High Energy Transient Explorer (HETE-2) mission is devoted to 
the study of GRBs using soft X-ray, medium X-ray, 
and gamma-ray instruments mounted on a compact spacecraft. 
HETE-2 was primarily developed
and fabricated in-house at the MIT by a small scientific and
engineering team, with major hardware and software contributions
from international partners in France and Japan \citep{doty}. Contributions to software development were also
made by scientific partners in the US, at the Los Alamos National
Laboratory, the University of Chicago, and the University
of California at Berkeley. Operation of the HETE satellite
and its science instruments, along with a dedicated tracking
and data telemetry network, is carried out by the HETE Science
Team itself \citep{crew}.
The spacecraft was successfully launched into equatorial orbit on 9~october~2000, and has operated in space during six years.
The GRB detection and localisation system on HETE consists
of three complementary instruments: the \textit{French Gamma
Telescope} (FREGATE), the \textit{Wide field X-ray Monitor} (WXM), and the \textit{Soft X-ray Camera} (SXC). The manner in which the three HETE science instruments
operate cooperatively is described in \citet{ricker}.

Since this study is based on the photon tagged data 
recorded by FREGATE, we now describe this instrument
(see \citet{atteia} for more details).
The main characteristics of FREGATE are given in Table~\ref{tab:fregate}.
The instrument, which was developed by the Centre d'Etude Spatiale
des Rayonnements (Toulouse, France), consists of four co-aligned 
cleaved NaI(Tl) scintillators, 
optimally sensitive in the 6 to 400~keV energy band, and one electronics box. 
Each detector has its own analog and digital electronics. 
The analog electronics contains a discriminator circuit with four adjustable
channels and a 14-bit PHA (\textit{Pulse Height Analyser}) whose output is regrouped into 256 evenly-spaced energy channels (approximately 0.8~keV or 3.2~keV wide). The (dead) time needed to encode the energy of each photon is 17 $\mu$s for the PHA and 9 $\mu$s for the discriminator. The digital electronics processes the individual pulses, to generate
the following data products for each detector: 

\begin{itemize}
\item Time histories in four energy channels, with a temporal resolution of 160~ms, 
\item 128-channel energy spectra spanning the range 6--400 keV, with a temporal resolution of 5.24~s,
\item A ``burst buffer'' containing the most recent 65\,536~photons 
tagged in time and in energy.
\end{itemize}

The burst buffers are only read when a trigger occurs. The four burst buffers
contain a total of 256k photons (64k per detector) tagged in time (with a resolution of 
6.4~$\mu$s) and in energy (256 energy channels). 
These data (also called photon tagged data) allow detailed studies of the
spectro-temporal evolution of bright GRBs. Given the small effective area 
of each detector (40~cm$^2$), the size of the burst buffers is usually sufficient
to record all the GRB photons. One exception is GRB~020813 which was made
of two main peaks separated by 60~s, and for which the burst buffers cover only the first peak.
This limitation does not affect the analysis presented here.

\section{Description of the analysis method}\label{sec:method}

The analysis of the 15 GRBs with measured redshifts follows the steps described below:

\begin{itemize}
\item
determination of the time interval to be studied between start and end of burst. It is defined by a cut above the background measured outside of the burst region,
\item
choice of the two energy bands for the time lag calculations, later called energy scenario, by assigning the individual photons to each energy band. The study of various scenarios is allowed by the use of tagged photon data provided by FREGATE for each GRB,
\item
de-noising of the light curves by a Discrete Wavelet Transform and pre-selection of data in the studied time interval for each GRB and each energy band,
\item
search for the rapid variations (spikes) in the light curves for all energy bands using a Continuous Wavelet Transform. The result of this step is a list of minima and maxima candidates, along with a coefficient characterising their regularity (Lipschitz coefficient $\alpha$ and its error $\delta\alpha$),
\item
association in pairs of the minima and of the maxima, which fulfill the conditions derived from studies of the Lipschitz coefficient.
\end{itemize}

As a result, a set of associated pairs is produced for each GRB and each energy scenario. The average time lag of each GRB, $<\Delta t>$, is then calculated and used later in the study of the Quantum Gravity model described in Section~\ref{sec:intro}. Finally, the evolution of the time lags as a function of $K_l$ variable allows us to constrain the minimal value of the Quantum Gravity scale $\mathrm{E}_\mathrm{QG}$.

\subsection{Use of Wavelet transforms in the analysis}

Wavelet analysis is being increasingly used in different fields like biology, computer science and physics \citep{dremin}. 

Unlike Fourier transform, wavelet analysis is well adapted to the study of non-stationary signals, \textit{i.e.} signals for which the frequency changes in time.

These methods will be applied to our data and illustrated through figures in Section~\ref{subsec:proc}.

\subsubsection{Wavelet Shrinkage}\label{subsubsec:waveshrink}

As explained by \citet{mallat}, the Wavelet Shrinkage is a simple and efficient method to remove the noise. The Discrete Wavelet Transform (DWT) is the decomposition of a signal at a given resolution level $L$ on an orthonormal wavelet basis. Such a basis can be defined by:
\begin{equation}
\label{eq:bonwa}
\left\lbrace\psi_{j,n} = \frac{1}{\sqrt{2^j}}\psi\left(\frac{t - 2^j n}{2^j}\right)\right\rbrace_{(j,n)\in\mathbb{Z}^2},
\end{equation}
where $\psi$ is the \textit{mother wavelet}. The decomposition provides some large coefficients corresponding to large variations (the signal), and small coefficients due to small variations (the noise). Then, applying a threshold to the wavelet coefficients and performing the inverse transform removes the noise from the signal. In the following, the wavelet \textit{Symmlet-10} is used.

There are different ways to apply a threshold to the coefficients.
In this study, we first shift the wavelet coefficients at fine scale so that their median value is set to unity and we apply the so-called \textit{soft thresholding} $\rho$ defined by:
\begin{equation}
\label{form:softthresh}
\rho_T (x) =
\left\lbrace
\begin{array}{ll}
x - T, & \mathrm{if}\ x \geq T \\
x + T, & \mathrm{if}\ x \leq -T \\
0, & \mathrm{if}\ |x| < T.
\end{array}
\right.
\end{equation}
where $x$ represents the wavelet coefficient and where $T$ is a given threshold.

The parameter $T$ is usually chosen so that there is a high probability to be above the maximum level of the noise coefficients. 
As proposed by \citet{donjohn}, we use the following relation:
\begin{equation}
\label{form:t}
T = \sigma \sqrt{2 \log N},
\end{equation}
where $\sigma$ is the noise and $N$ the number of bins of the signal.

It is possible to estimate the noise value $\sigma$ by using the wavelet coefficients at fine scale \citep{donjohn}.~After ordering the $N/2$ wavelet coefficients at fine scale, the median $M_X$ (rank $N/4$) is calculated. The estimator of $\sigma$ is then given by:
\begin{equation}
\label{form:estim}
\tilde{\sigma} = \frac{M_X}{0.6745}.
\end{equation}

\subsubsection{Wavelet Modulus Maxima}\label{subsubsec:mm}

The Continuous Wavelet Transform (CWT) can be used to measure the local variations of a signal and thus its local regularity. If a finite energy function $f$ is considered, its CWT is given by:
\begin{equation}
\label{form:cwt}
Wf(u,s) = \int_{-\infty}^{+\infty} f(t) \frac{1}{\sqrt{s}}\psi^{*}\left(\frac{t-u}{s}\right) dt,
\end{equation}
where $\psi^{*}$ is the wavelet function, scaled by $s$ and shifted by $u$. It is possible to demonstrate that there cannot be a singularity of the signal without an extremum of its wavelet transform. A \textit{modulus maxima line} is a set of points $(u, s)$ for which the modulus of the wavelet transform $|Wf(u,s)|$ is maximum.

It is, then, possible to detect with high precision an extremum in the data by looking at wavelet transform local maxima with decreasing scale, \textit{i.e.} in the region of zoom of signal details.

When using discrete signals, the minimum scale for looking at details should not be smaller than the width of one bin. In addition, the scale should not be larger than the width of the range in which the data is defined. So if the data is defined on the range $\lbrack 0,1 \rbrack$ by $N$ bins, one must have
\begin{equation}
\label{form:cond}
N^{-1} < s < 1.
\end{equation}

Here it is important to choose a wavelet for which modulus maxima lines are continuous when the scale decreases. This ensures that each extremum of the light curve is localised by only one continuous wavelet maxima line. In the following, the second derivative of a gaussian, known as \textit{Mexican Hat} wavelet is used.

Like in \citet{ellis}, extrema localised by the CWT are characterised by using the Lipschitz regularity, which is a measurement of the signal local fluctuations. The Lipschitz regularity condition is defined as follows: $f$ is pointwise Lipschitz $\alpha$ at $\nu$, if there exists a polynomial $p_{\nu}$ of maximum degree $\alpha \ge 0$ such that:
\begin{equation}
\label{form:lipschitz}
| f(t) - p_{\nu}(t)| \le \mathcal{K} |t-\nu|^\alpha,
\end{equation}
where $\mathcal{K}$ is a constant. This mathematical expression is not used as it is in our analysis, instead 
an estimation of the Lipschitz coefficient $\alpha$ is obtained from a study of the decrease of the wavelet coefficients along a maxima line at fine scales. For $s < s_0$, the following relation is used \citep{mallat}:
\begin{equation}
\label{form:decay}
\log_2 |Wf(u,s)| \approx \left(\alpha + \frac{1}{2}\right) \log_2 s + \mathrm{k},
\end{equation}
where k is a constant. $s_0$ is chosen so that it is smaller than the distance between two consecutive extrema. The values of the coefficient $\alpha$ and of its error $\delta\alpha$ are determined by a fit to Equation~\ref{form:decay}.

\subsection{Data sample: light curves}\label{subsec:dspresel}

From September 2001 to May 2006, FREGATE observed 15 GRBs for which redshift determination was possible and for which photon tagged data are available. These GRBs are located from $z = 0.16$ to $z = 3.37$. Table~\ref{tab:data} shows the redshift, corresponding value $K_l(z)$ (see Equation~\ref{form:kl}), duration and luminosity of each burst.

\begin{figure}
\epsscale{.90}
\plotone{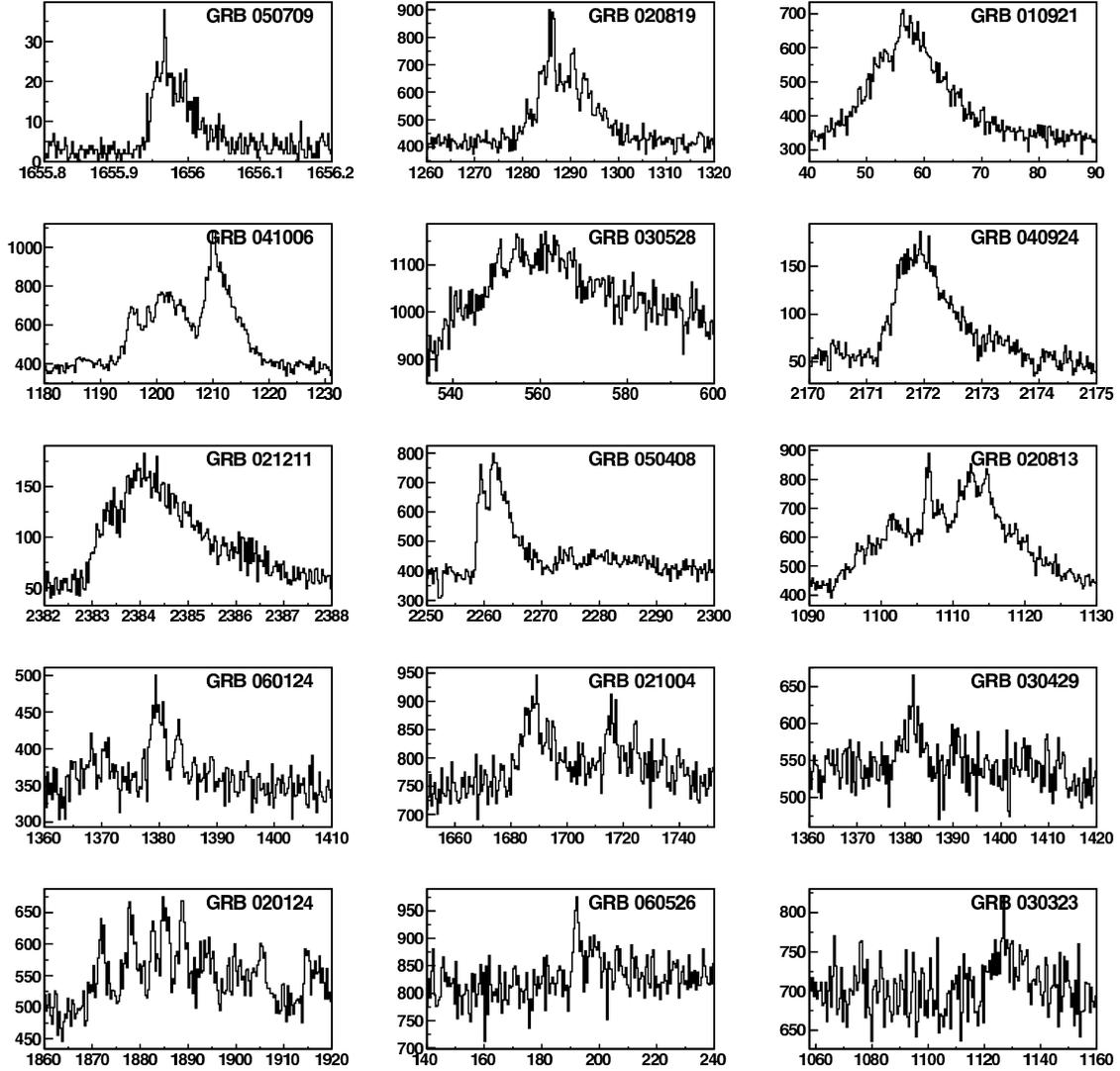}
\caption{Light curves of the 15 GRBs detected by HETE-2 in the energy range 6--400~keV. The bursts are sorted by increasing $z$ from top left to bottom right. X-axes are graduated in seconds.}
\label{fig:lcall}%
\end{figure}

The light curves of the 15 GRBs are shown in Figure~\ref{fig:lcall}. It appears clearly that the signal-to-noise ratio decreases for large redshifts. The large variety of light curve shapes is a characteristic feature of GRBs. Different binnings are chosen depending on the duration of the burst. For example, we use bins of 39~ms for GRB~020813 and 12~ms for GRB~021211. The pre-burst sections of the light curves are used to measure the background mean and variance $\sigma_\mathrm{bck}$.

\begin{figure}
\epsscale{.80}
\plotone{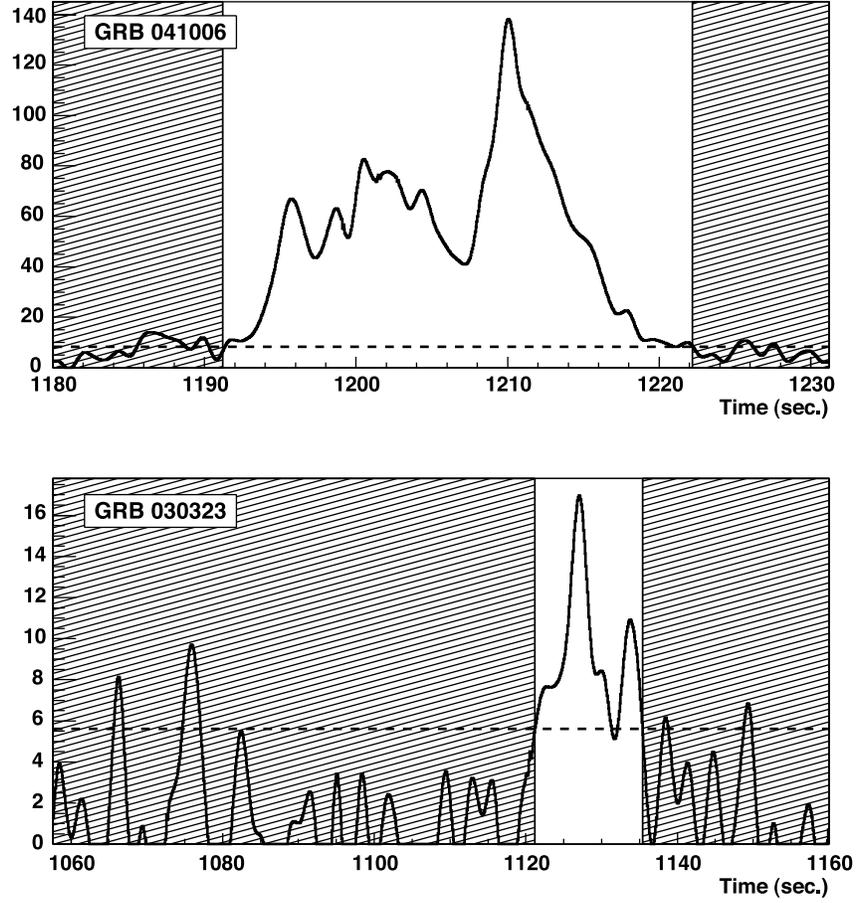}
\caption{Noise-free, background subtracted light curve of GRB~041006 (top) and of GRB~030323 (bottom). For GRB~041006, the horizontal line corresponds to $1\,\sigma_\mathrm{bck}$. For GRB~030323, the horizontal line corresponds to $0.5\,\sigma_\mathrm{bck}$. The hatched areas show the parts of the light curve that do not contribute to the analysis.}
  \label{fig:range1s}%
\end{figure}

After de-noising (described in Section~\ref{subsubsec:waveshrink}) and background removal, a time range in which the signal is greater than $\sigma_\mathrm{bck}$ is obtained (see top panel of Figure~\ref{fig:range1s}). For GRB~030323, GRB~030429 and GRB~060526, a level of $0.5\,\sigma_\mathrm{bck}$ is used because of a significantly lower signal-to-noise ratio (see bottom panel of Figure~\ref{fig:range1s}). In the following, only the part of the light curves located in this interval are considered for measurement of the time lags. Thus, most of the considered extrema have a source origin.

\subsection{Choice of energy bands}\label{subsec:ebands}

FREGATE measures the photon energies between 6 and 400~keV. After de-noising the light curves, extrema identification and pair association are done in energy bands. The selection of various energy bands is done by using the part of the spectra where the acceptance corrections are well understood and do not vary from burst to burst. Another important consideration concerns the maximisation of the energy difference between the low and high energy bands. Note that the choice of contiguous energy bands provides more extrema and pair candidates but a smaller energy lever arm. In the following, the choice of a pair of energy bands will be quoted as energy scenarios. Some scenarios will produce correlated results due to the overlap of their energy limits. The different choices of the energy scenarios are summarised in Table~\ref{tab:ene}.

\subsection{Procedure for pair association}\label{subsec:proc}

\begin{figure}
\epsscale{.90}
\plotone{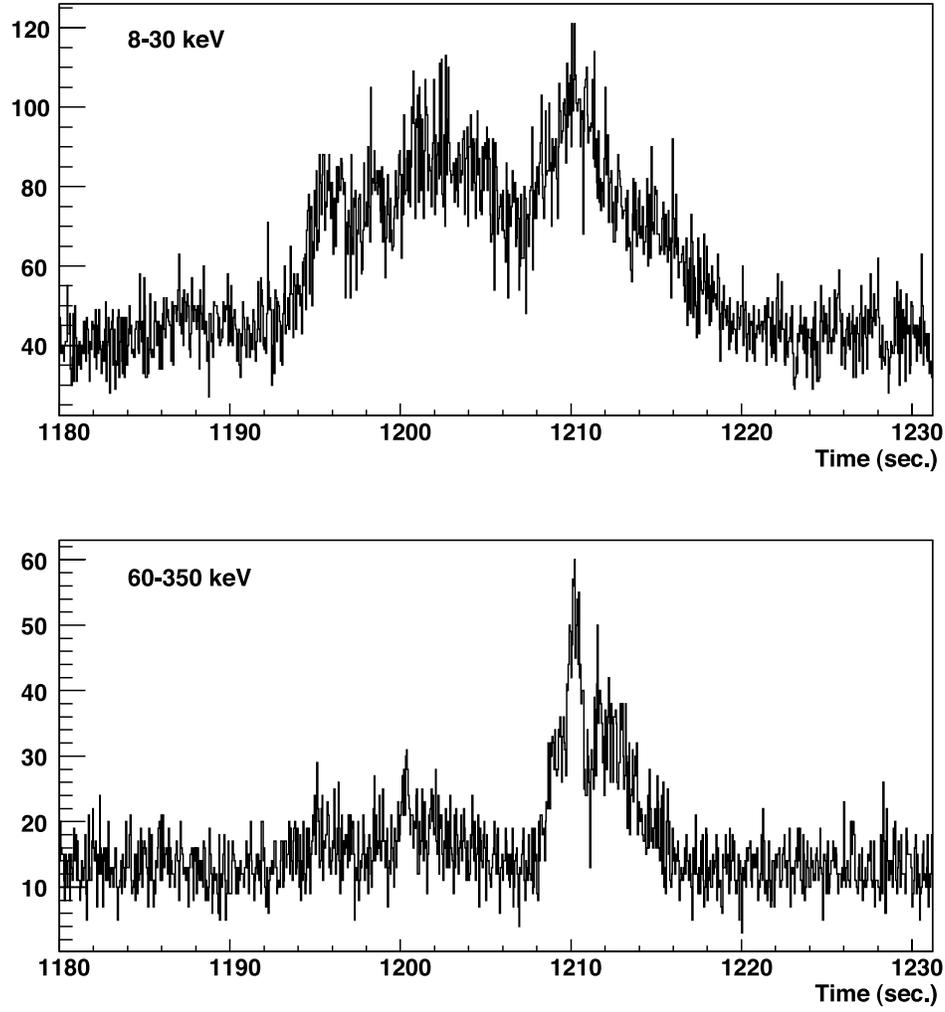}
\caption{Light curves of GRB~041006 for the energy bands 8--30 keV and 60--350 keV.}
\label{fig:lc}%
\end{figure}

\begin{figure}
\epsscale{.90}
\plotone{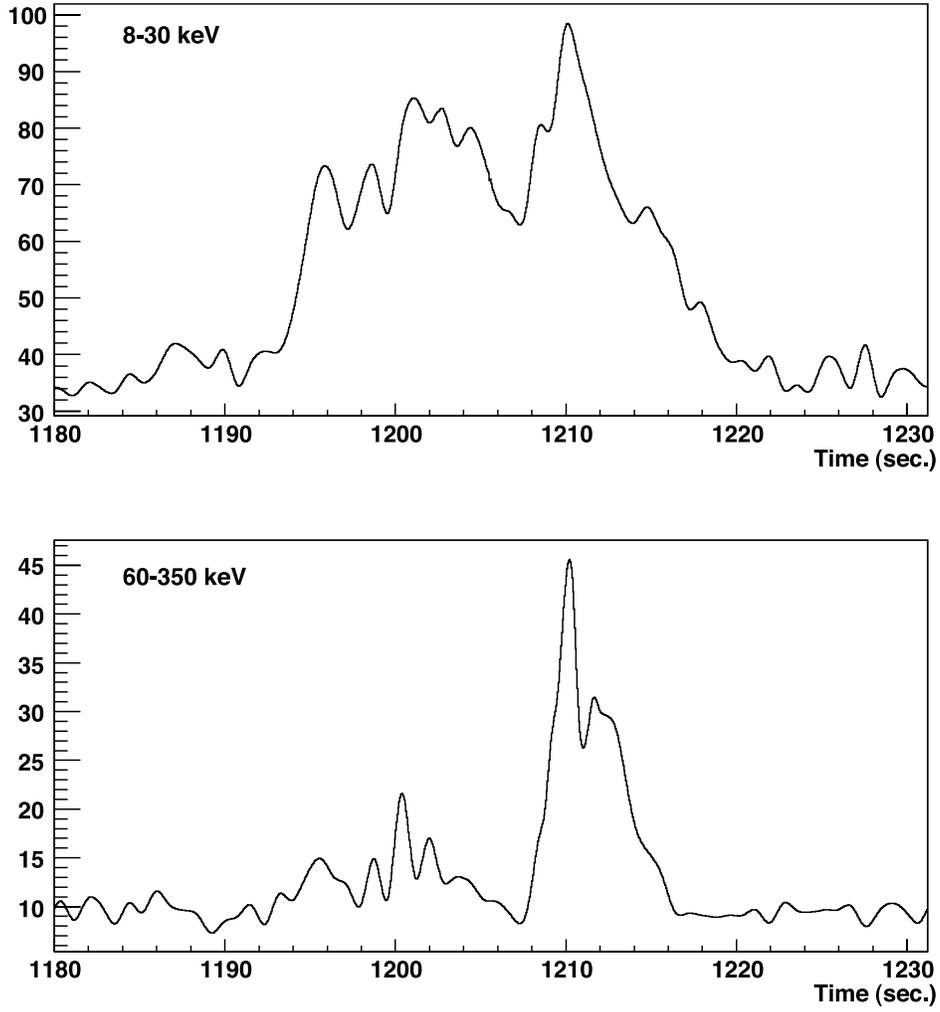}
\caption{De-noised light curves of GRB~041006 for the energy bands \mbox{8--30 keV} and \mbox{60--350 keV}. These curves were obtained with the wavelet Symmlet-10 with the soft thresholding proceedure at the level $L = 6$.}
\label{fig:noiserem}%
\end{figure}

\begin{figure}
\epsscale{.80}
\plotone{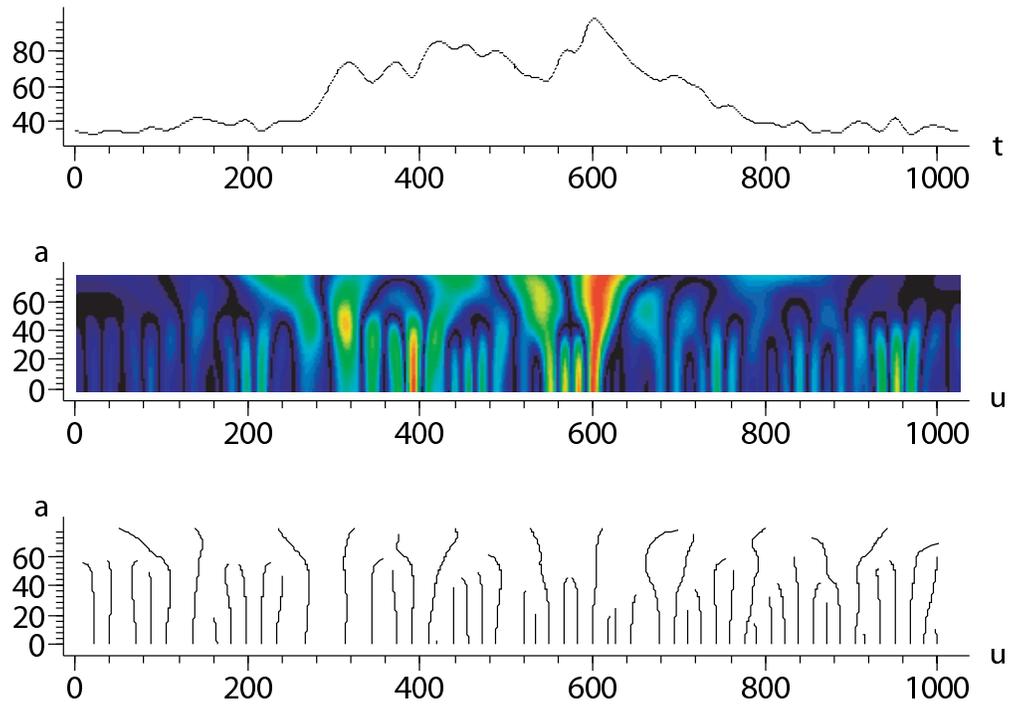}
\caption{The Continuous Wavelet Transform of the light curve between 8 and 30 keV of GRB~041006. The top panel shows the denoised data. The middle panel shows the wavelet transform, \textit{i.e.} the value of $|Wf(u,s)|$ for each $(u, s)$. The horizontal and vertical axes give respectively $u$ and $a = \log_2 s$. The bottom panel shows the modulus maxima of $Wf(u,s)$. Each maxima line corresponds to a singularity in the signal.}
\label{fig:mm}%
\end{figure}

Two light curves are obtained for two different energy bands, according to the method described in Section \ref{subsec:ebands} (see Figure~\ref{fig:lc}). Noise is removed \mbox{using} the MatLab package WaveLab \citep{wavelab} with the wavelet shrinkage procedure described in Section~\ref{subsubsec:waveshrink} (see~Figure~\ref{fig:noiserem}). Then the software LastWave \citep{lastwave} is used to compute the Continuous Wavelet Transform (see~Figure~\ref{fig:mm}). It gives a list of extrema for each light curve. 

Minima and maxima are sorted in two separate groups. As maxima correspond to a peak of photon emission and minima correspond to a lack of photons, they do not have the same behavior, \textit{a priori}.

As described in Section~\ref{subsubsec:mm}, the Lipschitz coefficient $\alpha$ and its error $\delta\alpha$ are deduced from wavelet coefficient decrease at fine scales. The derivative of the light curve $f(t)$ at the position of each extremum is also computed.

\begin{figure}
\epsscale{.80}
\plotone{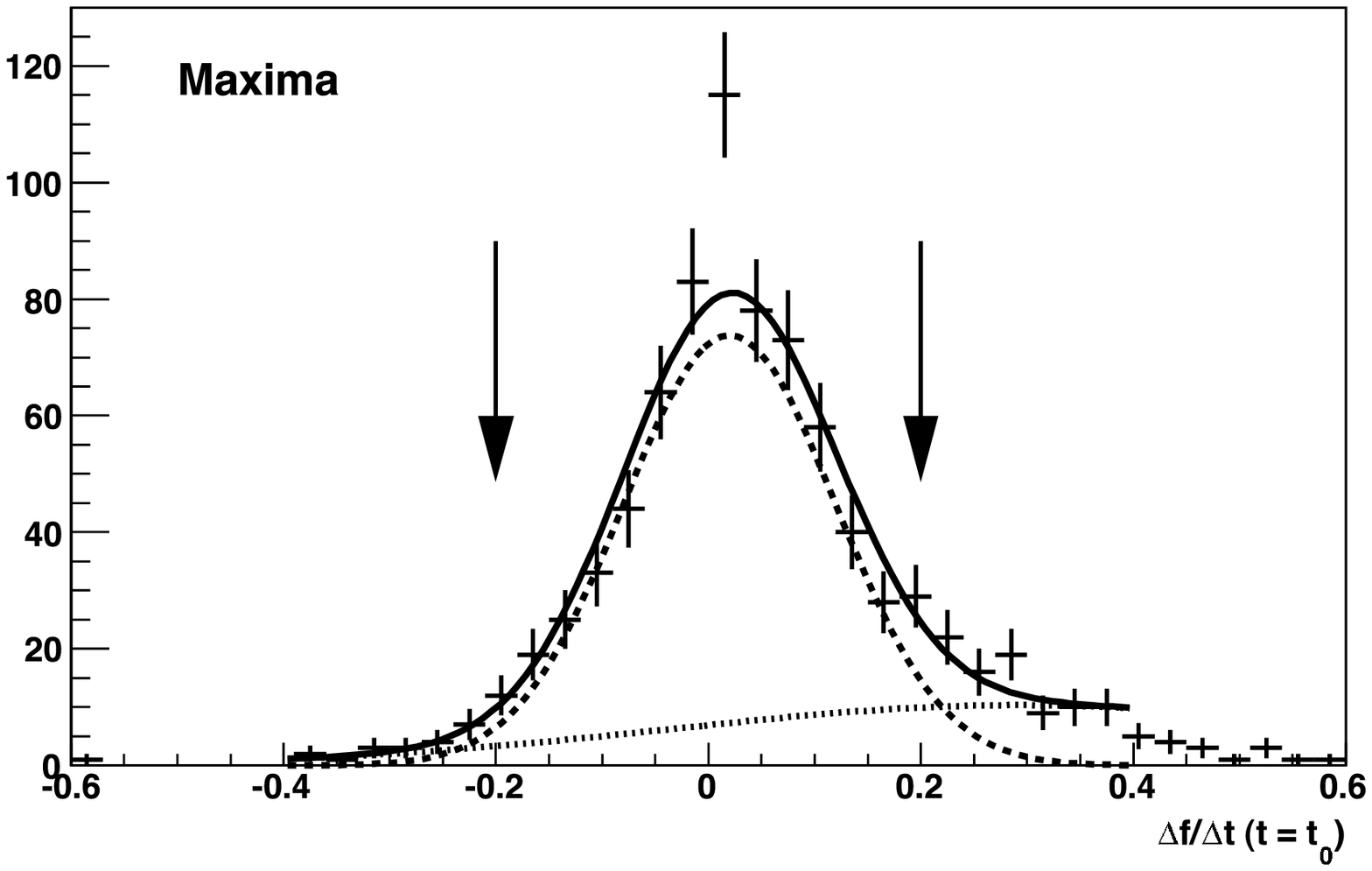}
\plotone{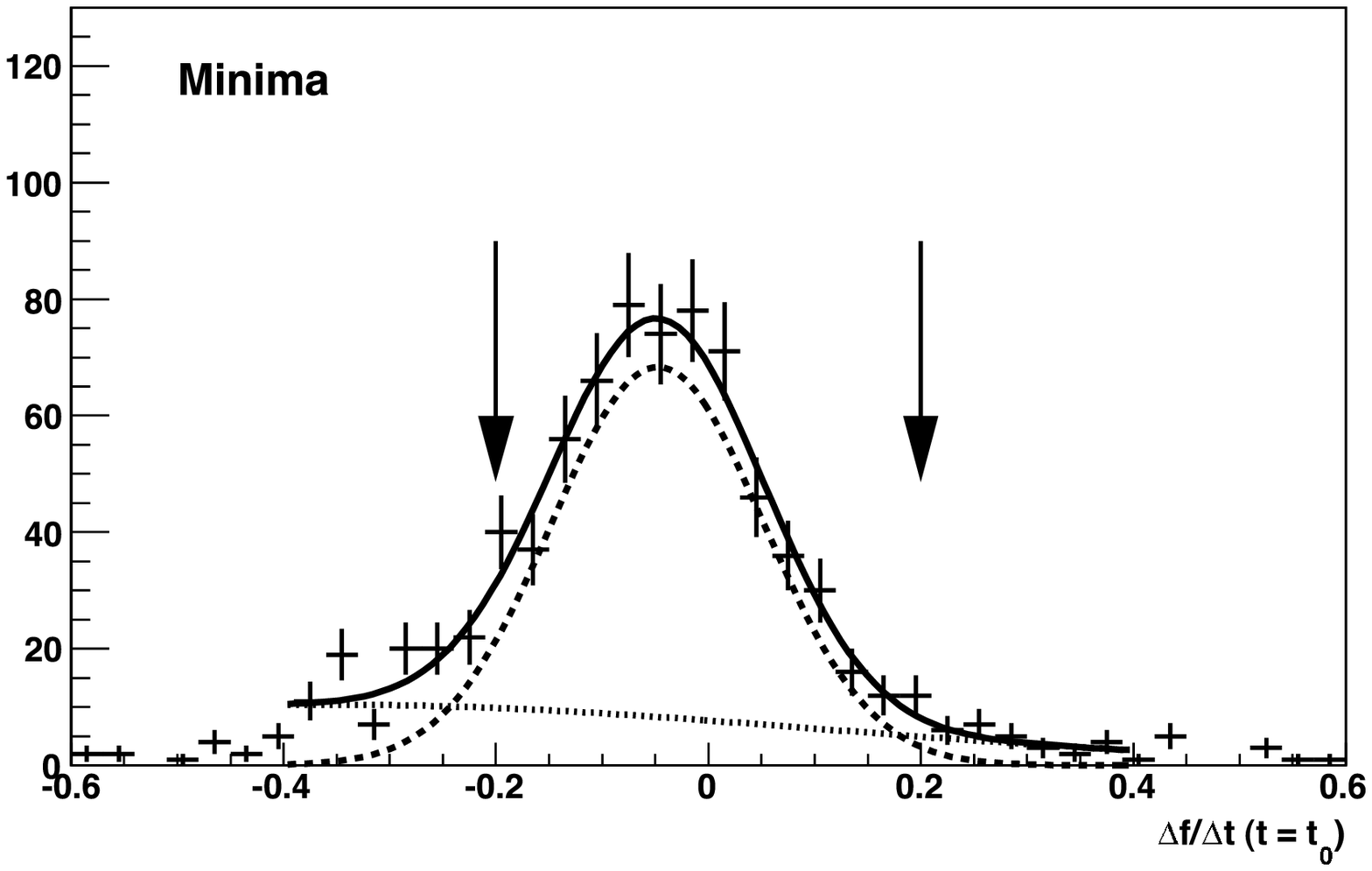}
\caption{Distribution of the derivatives of the light curve for maxima (top) and minima (bottom) located at $t_0$. Only extrema in the range found in Section~\protect\ref{subsec:dspresel} are considered. The data is fitted for \mbox{$-0.4 \leq (\Delta f/\Delta t)_{t = t_0} \leq 0.4$} with the sum of two Gaussian curves. The contribution of fake extrema is negligible for $|\Delta f/\Delta t| \leq 0.2$.}
\label{fig:distder2}%
\end{figure}

Figure~\ref{fig:distder2} shows that most of the extrema found by the CWT have non-zero derivatives. These curves can be described by a Gaussian curve centered at zero and a rather flat background centered on large positive (negative) values for the maxima (minima) candidates. 
The width of the Gaussian curve reflects measurement errors on extrema position determination.
The flat background corresponds to fake extrema found by our procedure.
Based on these distributions, the following cut was applied to ensure low contribution from fake extrema:
\begin{equation}
\label{form:cutder_sel}
\left|\frac{\Delta f}{\Delta t}\right| \leq 0.2.
\end{equation}
This condition preserves all 15 GRBs in our analysis.

A ``pair'' is made up of one extremum in the low energy band and one in the high energy band. Two values of time ($t_1$, $t_2$), two values of the Lipschitz coefficient ($\alpha_1$, $\alpha_2$) and two values of its error ($\delta\alpha_1$, $\delta\alpha_2$) are associated to each pair. Indices~1 and 2 are used for the low and high energy band, respectively.

To build a set of extrema pairs in the most unbiased way, three variables $\Delta t$, $\Delta\alpha$ and $\delta(\Delta\alpha)$ are used:
\begin{equation}
\label{form:cuts}
\left\lbrace
\begin{array}{l}
\Delta t = t_2 - t_1\\
\Delta\alpha = | \alpha_2 - \alpha_1 |\\
\delta(\Delta\alpha) = \sqrt{\delta\alpha_2^2 + \delta\alpha_1^2}. \\
\end{array}
\right.
\end{equation}

\begin{figure}
\epsscale{.90}
\plotone{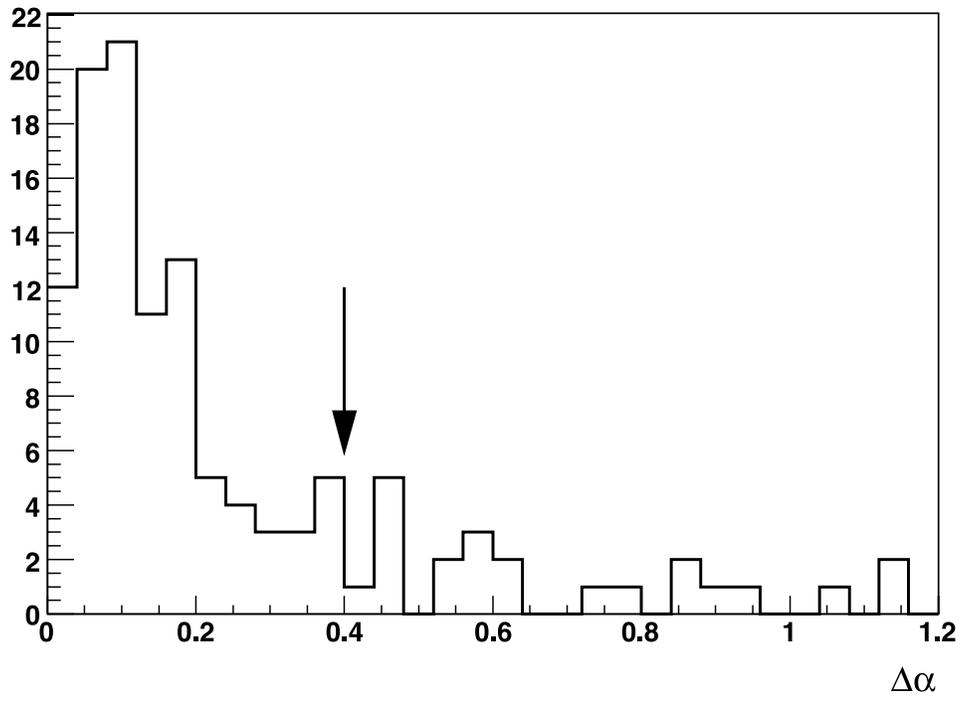}
\caption{Distribution of $\Delta\alpha$ for all candidate pairs in scenario~\#2. Cut is shown by an arrow.}
\label{fig:selalpha}%
\end{figure}

\begin{figure}
\epsscale{0.90}
\plotone{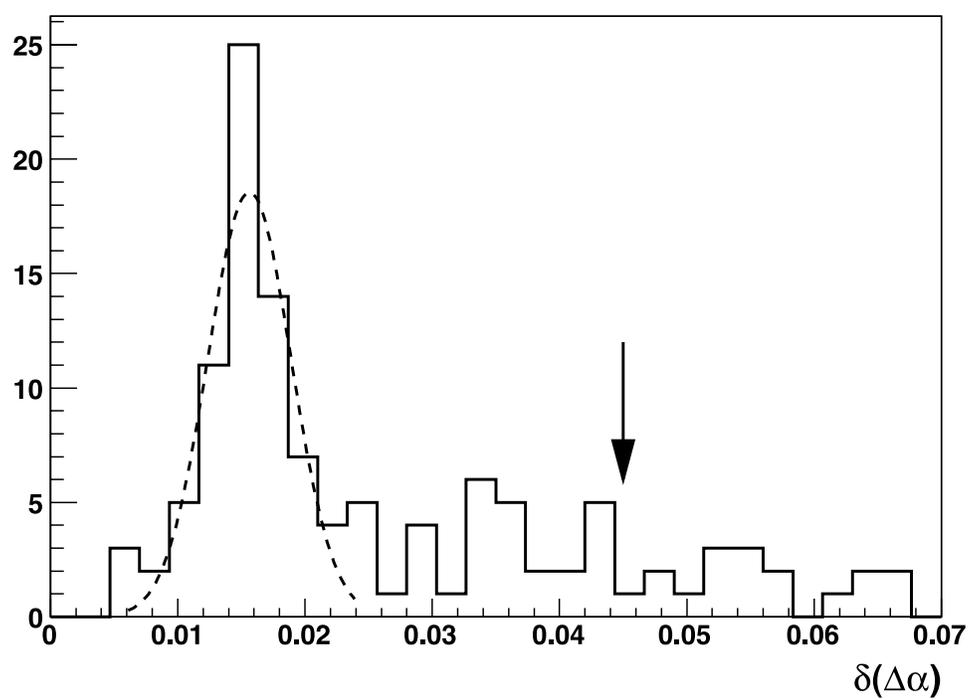}
\caption{Distribution of $\delta(\Delta\alpha)$ in the scenario~\#2. The cut is based on the position of the peak, determined by a Gaussian fit. We set the cut at three times the mean value of $\delta(\Delta\alpha)$.}
\label{fig:selsigma}%
\end{figure}

As we focus on QG effects which give only small time lags, we first select possible pairs with $| \Delta t | < 150\ \mathrm{ms}$ only. At this point each extrema can be used more than once depending on the distance between two consecutive irregularities. Then, based on the distributions of $\Delta\alpha$ and $\delta(\Delta\alpha)$ shown in Figures~\ref{fig:selalpha} and~\ref{fig:selsigma}, the following selections are applied:
\begin{equation}
\label{form:sels}
\left\lbrace
\begin{array}{l}
\Delta\alpha < 0.4 \\
\delta(\Delta\alpha) < 0.045. \\
\end{array}
\right.
\end{equation}

A small value of $\Delta\alpha$ ensures that the two associated extrema are of the same kind in the sense of the Lipschitz regularity. The cut on $\delta(\Delta\alpha)$ allows us to keep extrema mainly from the Gaussian peak in the $\delta\alpha$ distribution. These cut values are valid for all energy band choices.

After the selections, some extrema are used in more than one pair. To remove degeneracy in pair association, pairs which have the lowest $\Delta t$ are selected. This degeneracy concerns only $\sim6$\% of the total number of pairs before the cuts (Equation~\ref{form:sels}) are applied.

Table~\ref{table:nbpairs} shows the number of pairs found for all GRBs before and after all cuts. One may notice the stability of the cut efficiency leading to a mean value of $\sim$67\%.

\section{Results on Quantum Gravity scale}\label{sec:res}

\begin{figure}
\epsscale{0.60}
\plotone{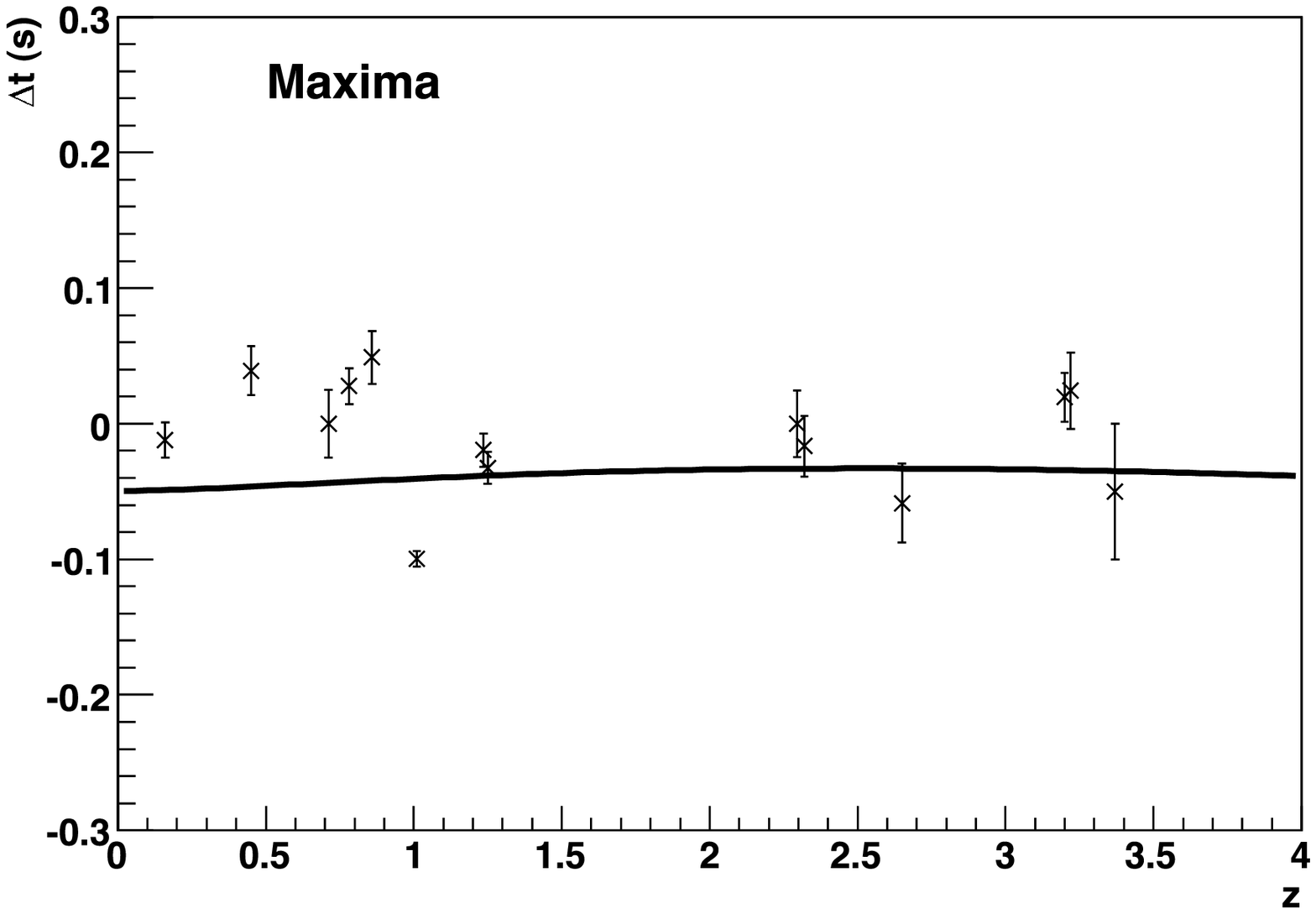}
\plotone{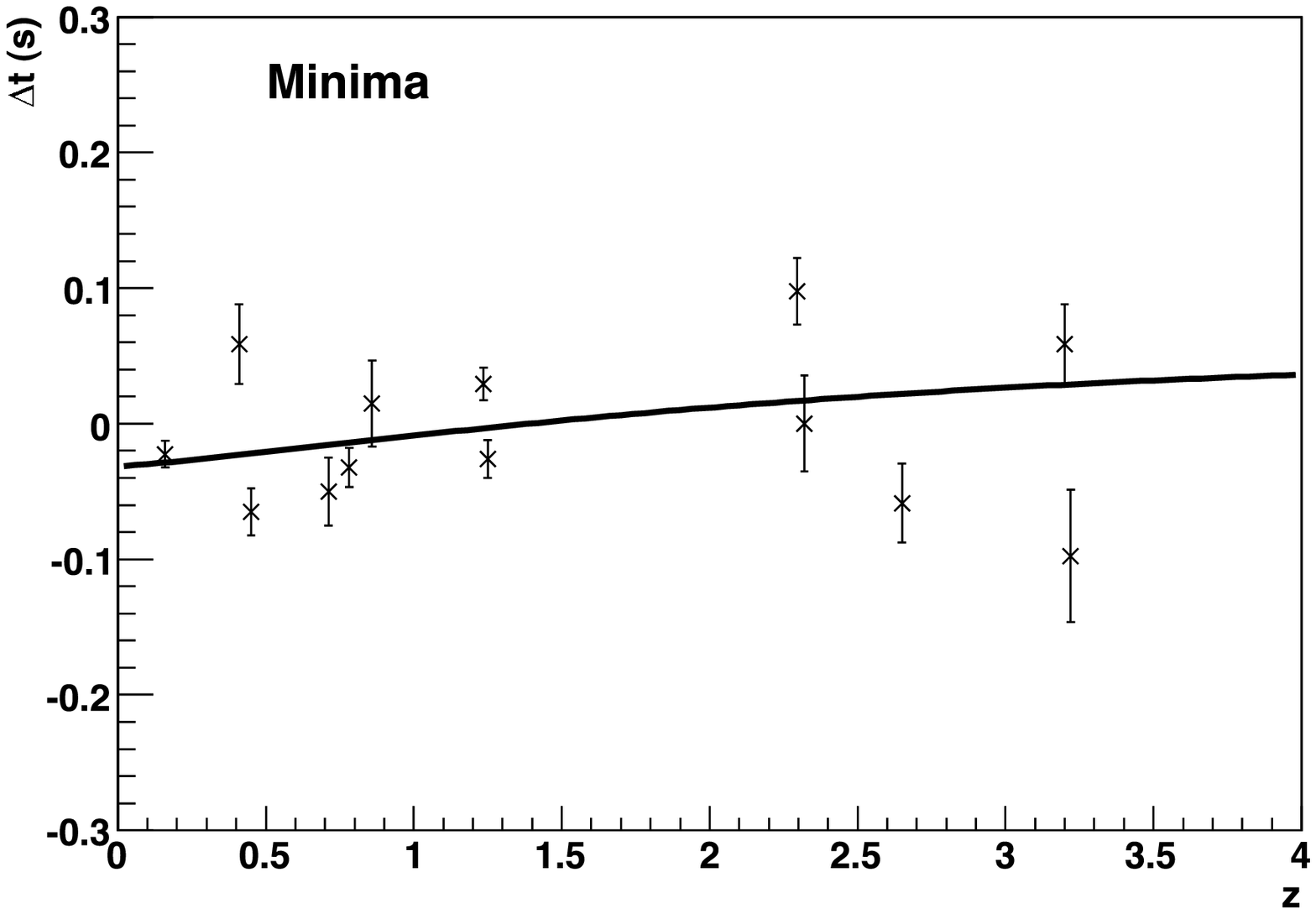}
\plotone{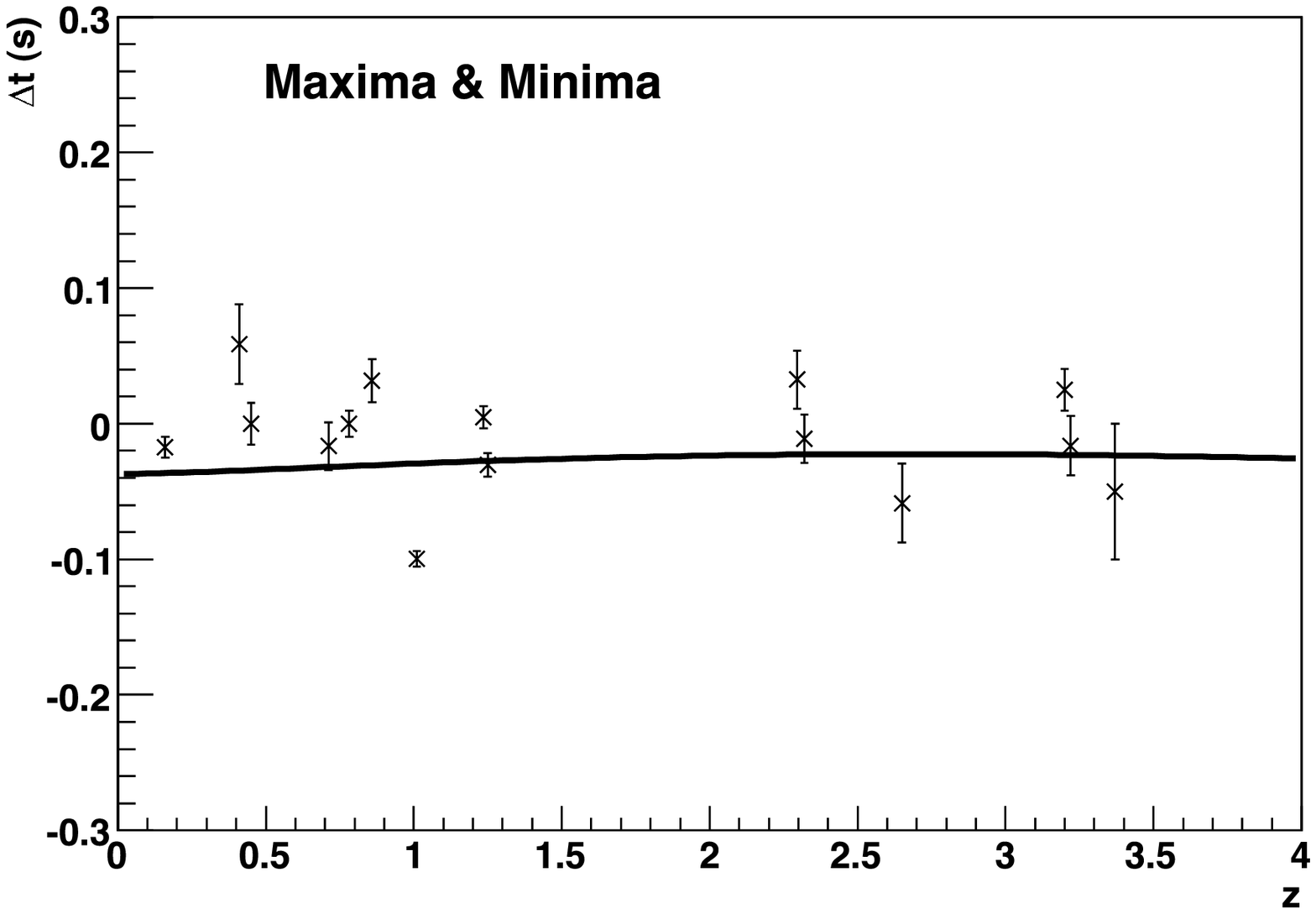}
\caption{$\Delta t$ in seconds, as a function of $z$ for scenario \#2, considering maxima only (top), minima only (middle) and both minima and maxima (bottom).
The curves show the result of a fit to Equation~\ref{form:linformab1}.}
\label{fig:resmaxmin}%
\end{figure}

\begin{figure}
\epsscale{0.60}
\plotone{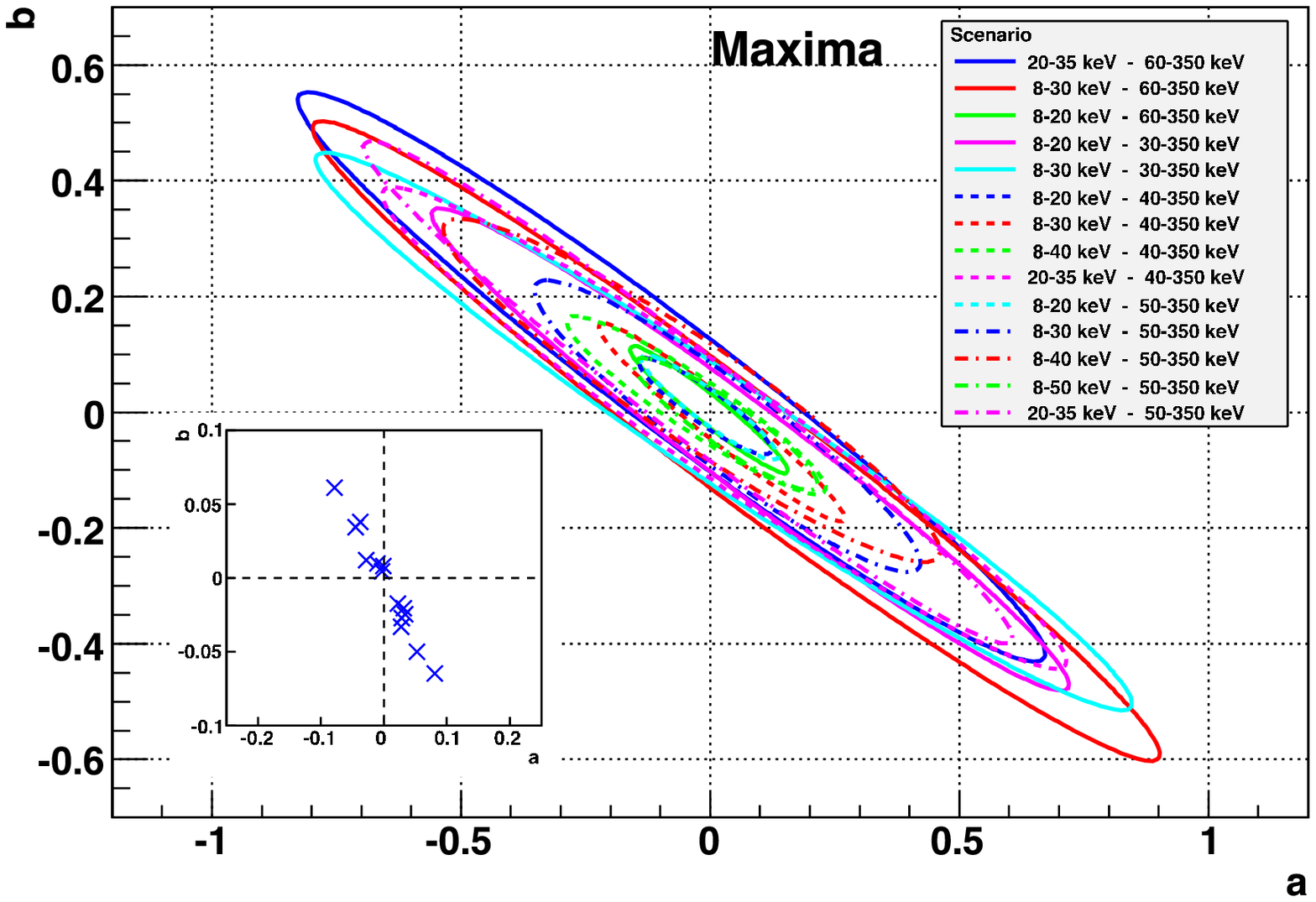}
\plotone{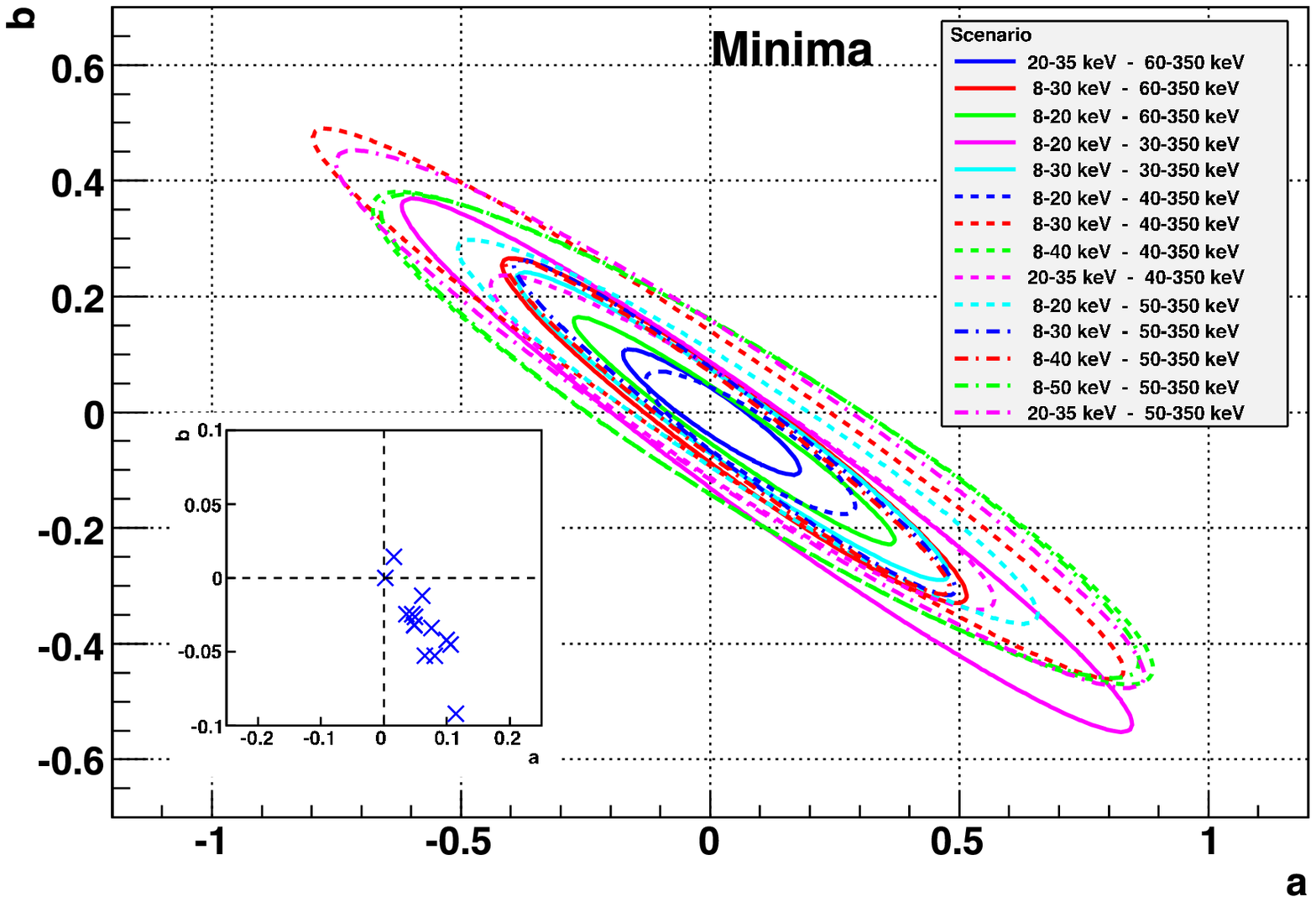}
\plotone{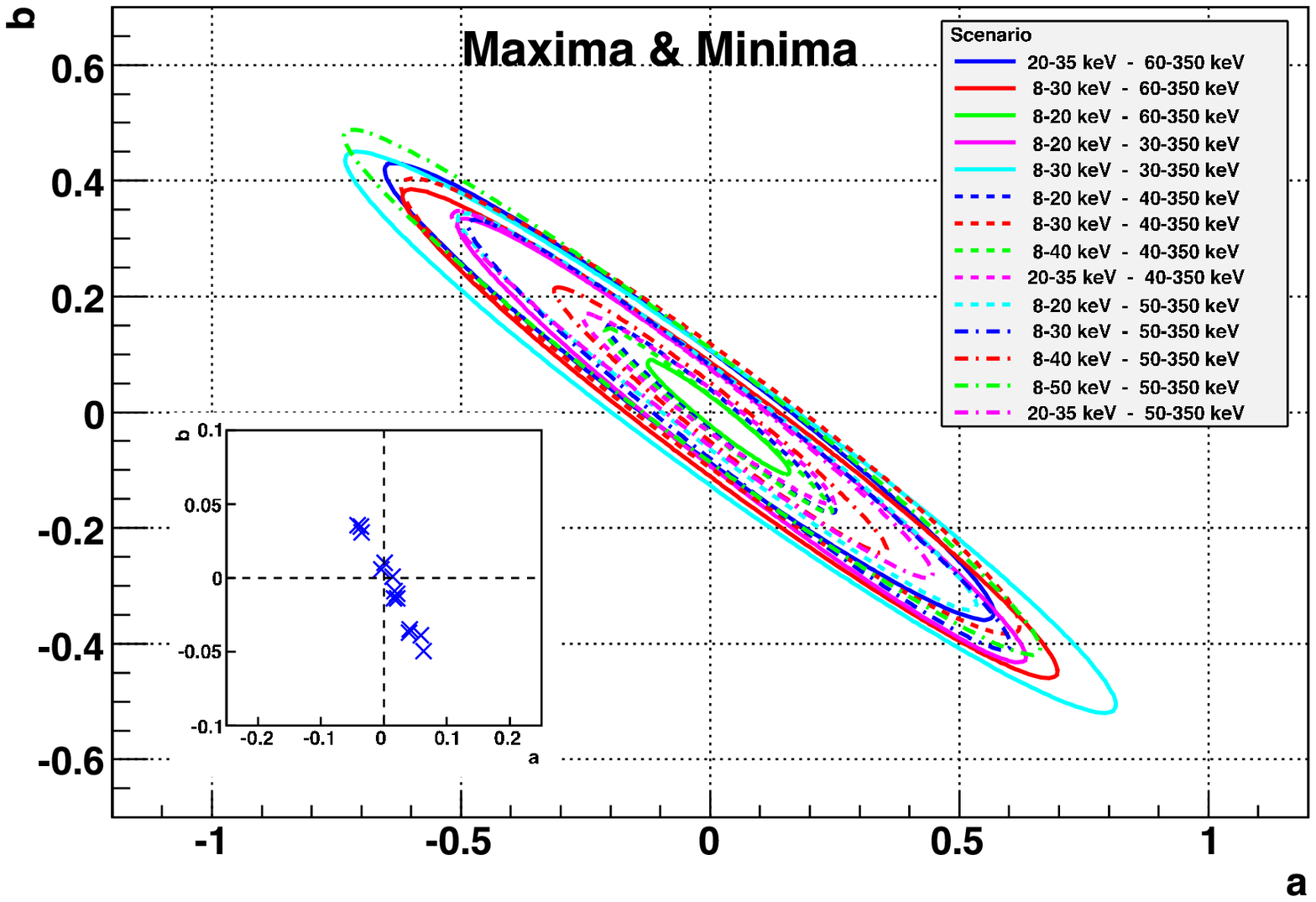}
\caption{95\% CL contours for $a$ and $b$ from the two-parameter fit for the fourteen scenarios, for maxima only (top), minima only (middle) and both minima and maxima (bottom). The boxes at the bottom left of the plots show the position of contour centers.}
\label{fig:avsb}%
\end{figure}

To study the model of Quantum Gravity described in Section~\ref{sec:intro}, the evolution of the mean time lags as a function of $z$ was determined using the set of 15 GRBs, as illustrated in Figure~\ref{fig:resmaxmin} for maxima, minima and combination of both for scenario \#2. We show on the same figure the fit of the data points with Equation~\ref{form:linformab1}. The results $(a,b)$ of the fits for all scenarios are summarised in Table~\ref{table:resab}. As explained in Section~\ref{sec:intro}, the parameter $a$ depends on the Quantum Gravity scale while the parameter $b$ reflects intrinsic source effects.
Both parameters were found to be strongly correlated, as shown in Figure~\ref{fig:avsb} which represents the 95\% Confidence Level (CL) contours for $a$ and $b$. In spite of the fact that maxima and minima present similar exclusion domains, it should be noticed that most ellipse centres are grouped around zero value for $a$ and $b$ in case of maxima and when both minima and maxima are considered, whereas they are slightly shifted towards values of $a>0$ and $b<0$ for the minima. However, the fit results suggest no variation above $\pm$3\,$\sigma$, so that in the following, we derive the 95\%~CL lower Quantum Gravity scale limit, assuming no signal is observed.

The dependence of the Quantum Gravity scale parameter $\mathrm{E}_\mathrm{QG}$ on $z$ may also be constrained by a direct study of the sensitivity of the 15 GRB data to the model proposed by~\citet{ellis3}. We have built a likelihood function following the formula:
\begin{equation}
\label{form:lkh}
L = \exp\left(-\frac{\chi^2(M)}{2}\right), 
\end{equation}
where $M$ is the energy and the $\chi^2(M)$ is expressed as: 
\begin{equation}
\label{form:chi2}
\chi^2(M) = \sum_{i = 1}^{\mathrm{N}_\mathrm{GRB}} \frac{(\Delta t_i - \tilde{b}\,(1+z_i) - a_i(M)\,{K_l}_i)^2}{\sigma_i^2+\sigma_{\tilde{b}}^2},
\end{equation}
where the index $i$ corresponds to each GRB and $\mathrm{N}_\mathrm{GRB} \leq 15$ is the total number of GRBs with at least one pair, for a given scenario.
The dependence of $a$ on $M$ as predicted by the considered model of Quantum Gravity is given by: 
\begin{equation}
\label{form:ai}
a_i(M) = \frac{1}{H_0}\,\frac{\Delta <E>_i}{M}.
\end{equation}
The mean values of energy are computed for each GRB and each scenario from the relation:
\begin{equation}
\label{form:de}
\Delta<E>\ =\ <E>_2 - <E>_1,
\end{equation}
where indices 1 and 2 represent the low and the high energy band, respectively. The averaged values of $\Delta<E>$ for all bursts are given in Table~\ref{tab:ene} for each scenario. In this study, a universality of the intrinsic source time lags has been assumed.
The average value $\tilde{b}$ (and its error $\sigma_{\tilde{b}}$) is obtained as the weighted mean of the values $b_k$ (and their errors $\sigma_k$) from the previous two-parameter linear fit:
\begin{equation}
\label{form:berrb}
\tilde{b} = \frac{\sum_k w_k\,b_k}{\sum_k w_k}\ \ \mathrm{and}\ \ \sigma_{\tilde{b}} = \frac{1}{\sqrt{\sum_k w_k}},
\end{equation}
where the index $k$ corresponds to each scenario and $w_k = 1/\sigma_k^2$.
Using the values of $b_k$ and $\sigma_k$ given in Table~\ref{table:resab}, one gets $\tilde{b} = 0.0023\pm0.0026$ for maxima, $\tilde{b} = -0.0282\pm0.0038$ for minima and $\tilde{b} = -0.0069\pm0.0035$ for both minima and maxima.

\begin{figure}
\epsscale{0.60}
\plotone{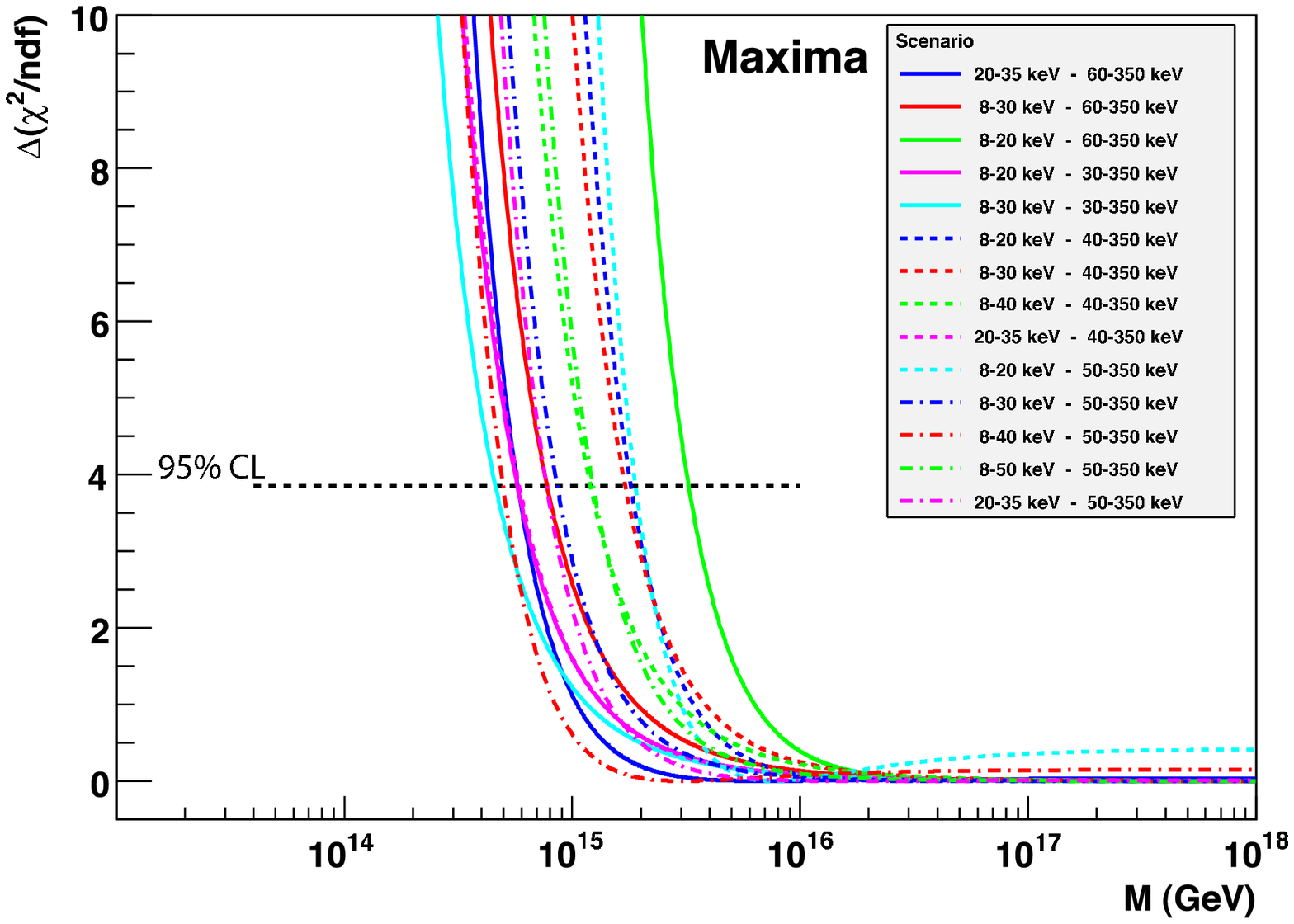}
\plotone{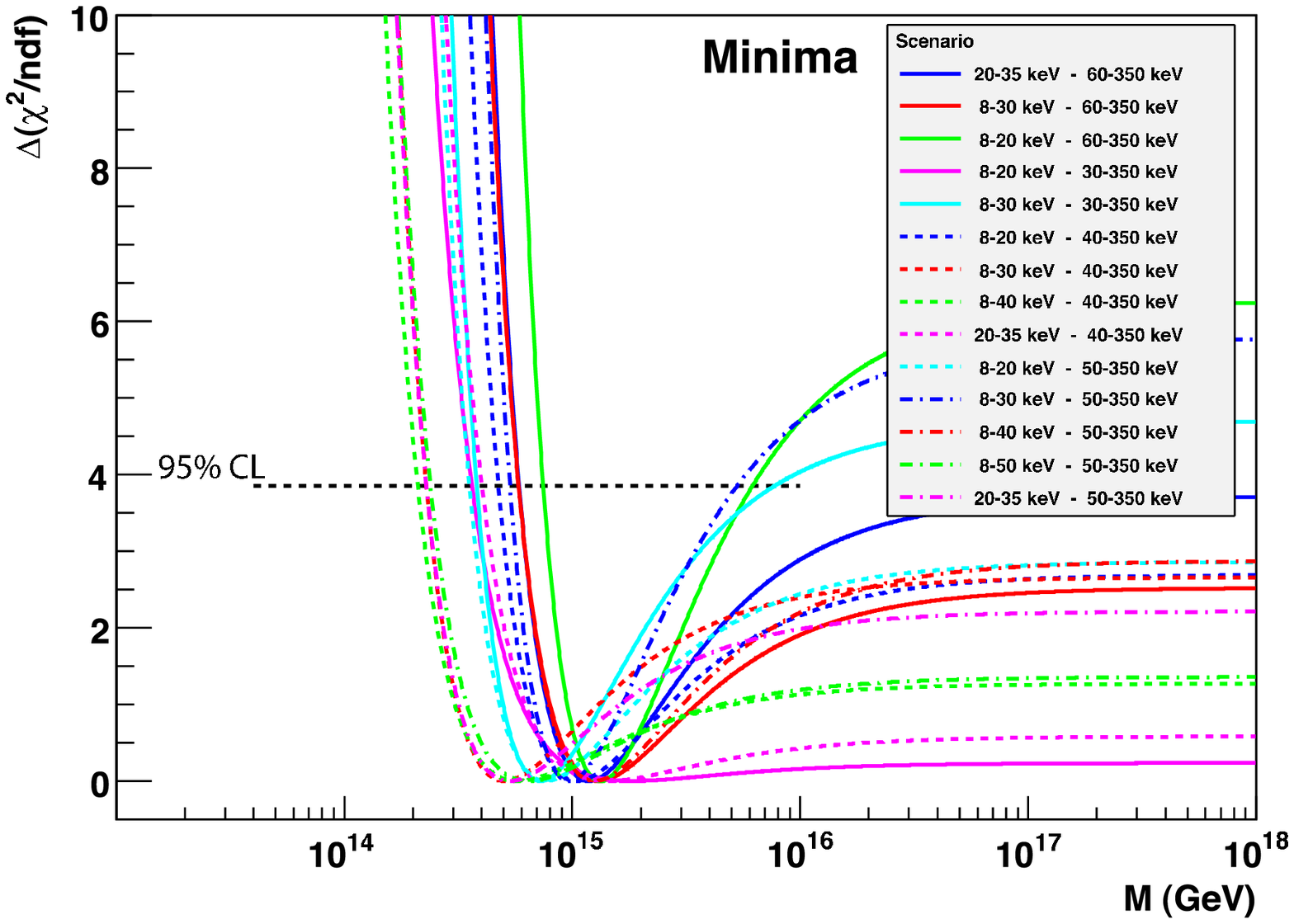}
\plotone{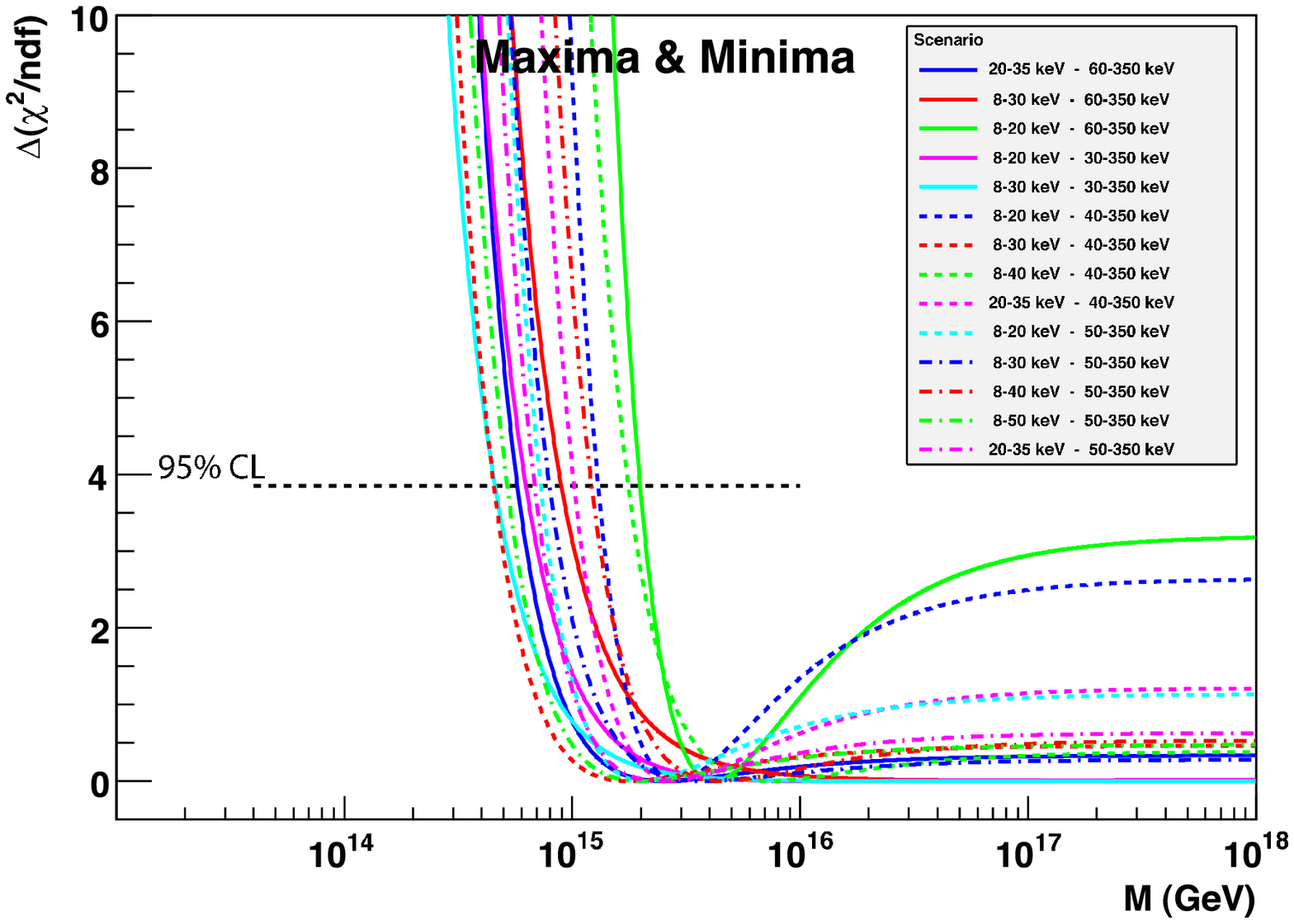}
\caption{Evolution of $\chi^2$ function of $M$, for maxima only (top), minima only (middle) and both minima and maxima (bottom).}
\label{fig:allqg}%
\end{figure}

Figure~\ref{fig:allqg} presents the evolution of $\chi^2(M)/\mathrm{ndf}$ around its minimum $\chi_{min}^2(M)/\mathrm{ndf}$ for maxima, minima and all extrema.
All scenarios fulfill the condition $\chi_{min}^2/\mathrm{ndf} \le 2$.
Like in the two-parameter linear fit, these curves show a different behaviour in case of the maxima and minima:
no significant preference of any value of $M$ is observed for most of the scenarios for the maxima, whereas a preferred minimum seems to be found for the minima. 

The 95\%~CL lower limit on the Quantum Gravity scale is set by requiring $\Delta(\chi^2/\mathrm{ndf})$ to vary by 3.84 from the minimum of the $\chi^2$ function.
The values of limits on $\mathrm{E}_\mathrm{QG}$ (shown in Table~\ref{table:reseqg}) are obtained for the 14 energy scenarios. In good agreement with the slope parameter $a$, all values for minima and maxima are within the $10^{14}$-$10^{15}$~GeV range.
No correlation with energy band choice or with any other cut parameter is found.
As there is no reason to choose any particular value of the lower limit on the Quantum Gravity scale obtained for the 14 scenarios, we choose scenario \#3, which provides the strongest constraint on the QG scale. This gives us a lower limit on $\mathrm{E}_\mathrm{QG}$ of $3.2\times10^{15}$~GeV for the maxima and a smaller value of $7.5\times10^{14}$~GeV for the minima. Combining maxima and minima, the best lower limit for the QG scale remains of $2\times10^{15}$~GeV.

\section{Discussion on various origins of time lags}\label{sec:syseffect}

\subsection{Measurement uncertainties in the time lag determination}

The time lag determination procedure proposed in this analysis is subject to various systematic effects due to: 
\begin{itemize}
\item the experimental precision on time measurement and energy estimation by the FREGATE detector, 
\item the biases that could be introduced by the use of the wavelet methods, 
\item the procedures employed for the extrema selection and pair association.
\end{itemize}

The precision on the individual photon time measurement is governed by the precision on the HETE-2 internal clock time synchronised to UTC by a spatial GPS whithin few $\mu$s, resulting in an overall precision value below 1 ms, well below the bin width in the CWT decomposition. The energy resolution of the FREGATE detector in the studied domain does not exceed 20\% and has little impact on the results of the present analysis. The error on redshift determination by the optical observations of the GRBs, can be considered negligible as well.

The systematic effects of the wavelet decomposition may have impact not only on the de-noising of the light curves but also on the extrema localisation, producing different biases in the maxima and minima identification. This could be the explanation of the different behavior of $\chi^2$ curves for maxima and minima. Since systematic effects may be different for maxima and minima, we consider the result obtained by the combination of both maxima and minima as more reliable.

The result of the de-noising procedure depends on two parameters: the type of the wavelet function and the level of the decomposition $L$. Wavelet functions Symmlet-10 and Daubechies-10 were considered. Both have ten vanishing moments but Symmlet is more symmetric. Both wavelets give comparable results: no discrepancy was observed between extrema positions. Following the approach by \citet{ellis}, the Symmlet-10 wavelet is used for the results given in this paper.
As far as the decomposition level $L$ is concerned, different values were tested. The lower the value of $L$ is, the smoother is the obtained light curve and the fewer extrema are found. The value $L = 6$ was chosen because it allows us to keep a significant level of detail without adding too much noise. The noise fluctuation cut at 1$\sigma$ above background was also raised to higher values, consequently decreasing the number of extrema candidates. However, no significant change in the results was observed. 
The cuts applied on values of the derivative at each extrema found with the CWT were studied. A more stringent cut of $\leq 0.1$ value was also applied to reject the fake extrema. This cut rejects about 40\% of pairs giving less significant results, while the cut value of $0.2$ (Equation~\ref{form:cutder_sel}) rejects about 15\% of pairs. Concerning the cuts on $\Delta\alpha$ and $\delta(\Delta\alpha)$, two selections, more severe than Equation~\ref{form:sels}, have been investigated:
\begin{equation}
\label{form:sels2}
\left\lbrace
\begin{array}{l}
\Delta\alpha < 0.2 \\
\delta(\Delta\alpha) < 0.045 \\
\end{array}
\right.
\mathrm{\ and\ }
\left\lbrace
\begin{array}{l}
\Delta\alpha < 0.4 \\
\delta(\Delta\alpha) < 0.02. \\
\end{array}
\right.
\end{equation}

The different choices of cuts on the extrema selections and pair association had little effect on the final limits, even if the statistical sensitivity of the studied sample was reduced. In particular, fewer extrema candidates were found in the light curves, leading to a smaller GRB sample.

\subsection{Intrinsic source effects}

Even in the restricted range of energies in our study, the GRB light-curves are not ``perfect'' signals, since they exhibit intrinsic time lags between high and low energies, which vary from burst to burst. It has been known for a long time, that the peaks of the emission are shorter and arrive earlier at higher energies \citep[and references therein]{fenimore,norris6}. These intrinsic lags, which have a sign opposite to the sign expected from Lorentz violation, have a broad dispersion of durations, complicating the detection of Lorentz violation effects. Therefore the effect of Lorentz violation must be searched with a statistical study analysing the average dependence of the lags with $z$, and not with a single GRB.

In particular, a strong anticorrelation of spectral lags with luminosity has been found by \citet{norris}. At low redshifts, we detect bright and faint GRBs which present a broad distribution of instrinsic lags whereas at high redshifts, only bright GRBs with small intrinsic lags are detected. This effect could mimick the effect of Lorentz violation due to the non uniform distribution in luminosity of the GRBs in our sample.

Following \citet{ellis5}, we use the generic term of ``source effects'' for the intrinsic lags discussed above. Source effects must be carefully taken into account if one wants to derive meaningful limits on the magnitude of Lorentz violation. Indeed, the source effects mentioned afore could explain the positive signal recently reported by \citet{ellis5}. There are at least two ways to take into account the source effects:
\begin{itemize}
  \item Their modelisation, which allows to understand the impact of the sample selection on the final result.
  \item The selection of a GRB sample homogeneous in luminosity, which minimises the impact of source effects.
\end{itemize}

The modelisation of the source effects is beyond the scope of this paper (but it will become more and more important in future studies, when the increasing number of GRBs will allow to place stronger constraints on Quantum Gravity effects). Here, with limited statistics, we made a test by performing our analysis on a restricted GRB sample, almost homogeneous in luminosity.
The HETE-2 sample we use comprises 15 GRBs with redshifts ranging between 0.1 and 3.4, biased with respect to the luminosity population. The high redshift part of the sample is mainly populated by the high luminosity GRBs for which the source time lag effects are expected to be the weakest. The independent analysis of a GRB sample with luminosity values $\mathrm{L}_{51} > 8\:\mathrm{erg}\:\mathrm{s}^{-1}$ (10 GRBs) provided similar results to those obtained with full sample of 15 GRBs. The sensitivity was lower because of a decreased statistical power of the restricted sample and a smaller lever arm in the redshift values.

\section{Summary}\label{sec:summary}

We have presented in this paper the analysis of the time structure of 15 GRB light curves collected by the HETE-2 mission in years 2001--2006, for which the redshift values have been measured. This sample was used for time lag measurements as a function of redshift. These time lags can originate either from the astrophysical sources themselves or from a possible Quantum Gravity signature. The latter case would result in energy and redshift dependence of the arrival time of photons.

We have used one of the most precise methods based on the wavelet transforms for the de-noising of the light curves and for the localisation of sharp transitions in various energy ranges. 
The maxima and the minima in the light curves have been studied separately as well as together. In particular, the $\chi^2$ dependence on the Quantum Gravity scale for the maxima shows a continuous decrease with energy scale parameter $M$ for most of the energy scenarios, whereas a minimum value may be detected for the minima. For 14 choices of the energy bands, the observed slope as a function of redshift for the maxima does not exceed 3\,$\sigma$ variation from a zero value. The preferred value in case of the minima is situated between $10^{14}$ and $10^{15}$~GeV for almost all considered energy scenarios. We use minima and maxima together to determine the 95\% CL lower limit on the Quantum Gravity scale for two reasons. The first reason is that we cannot exclude zero values at 95\% CL for a and b parameters of the two parameter fit reflecting QG and source effects. The second is that there is only a slight difference between the results for minima and maxima. In most of the studied energy scenarios, the lower limit value is of the order of $10^{15}$~GeV and can be considered as competitive considering the modest energy gap of $\sim$130 keV provided by the HETE-2 data. In this study, we do not correct for the astrophysical source effects, except in case of the analysis performed on a restricted GRB sample, homogeneous in luminosity. 

The impact of the selections and cuts on the obtained results has also been investigated. The most important contribution to the systematic effects on the background suppression comes from the decomposition level choice in the discrete wavelet transformation used in the de-noising procedure. All cuts applied in the extrema identification and in the pair association have also been varied, and lead to compatible results with those obtained from the optimised selections. The other important factor in this analysis is related to the energy lever arm in the selected energy scenarios. Here, we would like to underline the importance of the access to the full information on energy and time variables provided by the FREGATE detector, delivered on an individual photon basis. The FREGATE precision on photon time measurement is also a crucial parameter for the proposed analysis.

In summary, studies of time lags from the position of all the extrema in the light curves of the HETE-2 GRBs allow us to set a lower limit on the Quantum Gravity scale of
$$\mathrm{E}_\mathrm{QG} > 2\times 10^{15}\ \mathrm{GeV}.$$
The $\chi^2$ curves for the minima of the light curves show a prefered minimum between $10^{14}$ and $10^{15}$~GeV. All studies of the systematic effects do not change these results in a significant way. Further improvement of the limits on the Quantum Gravity energy scale would need an order of magnitude larger statistics and a larger energy lever arm. The next step in this direction, would be the analysis of an enlarged sample of HETE-2 GRBs, with measured pseudo-redshift values \citep{pelan}.

\acknowledgments
The authors would like to thank the referee for a crucial remark about the basic formalism of this study and A. Blanchard for his help in the understanding of the cosmology scheme related to this kind of analysis.
We would also like to thank our colleagues from LPTA, E.~Buffenoir, G.~Moultaka, D.~Polarski and Ph.~Roche, for the enlightening discussions on Quantum Gravity foundations.
The authors acknowledge the work of the \mbox{HETE-2} operation team which permitted the detection and fast localisation of the GRBs studied in this paper. 
HETE is an international mission of the NASA Explorer program, run by the Massachusetts Institute of Technology. The HETE mission is supported in the U.S. by NASA contract NASW-4690; in Japan, in part by the Ministry of Education, Culture, Sports, Science, and Technology Grant-in-Aid 13440063; and in France, by CNES contract 793-01-8479. 
The work of J.~Bolmont was partly supported by the R\'egion Languedoc-Roussillon.

\appendix

\section{Time lags and cosmological effects}\label{an}

The relation between time and redshift is given by
\begin{equation}
\label{form:time_z}
dt = -\mathrm{H}_0^{-1}\ \frac{dz}{(1 + z)\,h(z)},
\end{equation}
where $h(z)$ is defined by Equation~\ref{form:hz} and where $\mathrm{H}_{0}$ is the Hubble constant.

During a time $dt$, a particle with velocity $u$ travels a distance of:
\begin{equation}
\label{form:an_dl}
dl = u\,dt = -\mathrm{H}_0^{-1}\ \frac{u\,dz}{(1 + z)\,h(z)}.
\end{equation}
It is important to note here that this distance is measured at redshift $z$. The same distance measured at the redshift of the oberver is then given by
\begin{equation}
dl_0 = -\mathrm{H}_0^{-1}\ \frac{u\,dz}{h(z)}.
\end{equation}

So, two particles with velocities different by $\Delta u$ travel different distances by:
\begin{equation}
\label{form:DL}
\Delta L = \mathrm{H}_0^{-1} \frac{\Delta u\,dz}{h(z)}.
\end{equation}

From Equation \ref{form:di} (Section~\ref{sec:intro}), we deduce that two photons with an energy difference $\Delta E$ at redshift $0$ present a velocity difference at redshift $z$ of:
\begin{equation}
\Delta u = \mathrm{-c}\ \frac{\Delta E\,(1 + z)}{\mathrm{E}_\mathrm{QG}}.
\end{equation}

The integration of equation \ref{form:DL} provides the final formula (Equation~~\ref{form:dtc} in the text) for the time lag when the two photons are emitted at the same time:
\begin{equation}
\Delta t = \mathrm{H}_0^{-1} {\frac{\Delta E}{\mathrm{E}_\mathrm{QG}}} \int_{0}^{z} \frac{(1 + z)\,dz}{h(z)}.
\end{equation}


\clearpage

\begin{deluxetable}{lrr}
\tabletypesize{\scriptsize}
\tablecaption{Main parameters of the FREGATE detector.\label{tab:fregate}}
\tablewidth{0pt}
\tablehead{
\colhead{Parameter} & \colhead{Value}
}
\startdata
Energy range                             & 6--400 keV \\
Effective area (4 detectors, on axis)    & 160 cm$^2$ \\
Field of view (HWZM)                     & 70$\degr$ \\
Sensitivity (50--300 keV)                & 10$^{-7}$ erg cm$^{-2}$ s$^{-1}$ \\
Dead time                                & 10 $\mu$s \\
Time resolution                          & 6.4 $\mu$s \\
Maximum acceptable photon flux           & 10$^3$ ph cm$^{-2}$ s$^{-1}$\\
Spectral resolution at 662 keV           & $\sim$8\%  \\
Spectral resolution at 122 keV           & $\sim$12\%  \\
Spectral resolution at 6 keV             & $\sim$42\%  \\
\enddata
\end{deluxetable}

\begin{deluxetable}{ccccc}
\tabletypesize{\scriptsize}
\tablecaption{15 GRBs observed by HETE-2 from September 2001 to May 2006.\label{tab:data}}
\tablewidth{0pt}
\tablehead{
\colhead{GRB} & \colhead{z} & \colhead{$K_l$} & \colhead{T90 (s)} & \colhead{$\mathrm{L}_{51}$ ($\mathrm{erg}\:\mathrm{s}^{-1}$)}
}
\startdata
GRB~050709 & 0.16 & 0.17 & 0.1  & -     \\
GRB~020819 & 0.41 & 0.44 & 12.6 & 0.64  \\
GRB~010921 & 0.45 & 0.49 & 21.1 & 1.31  \\
GRB~041006 & 0.71 & 0.79 & 19.0 & 5.46  \\
GRB~030528 & 0.78 & 0.87 & 21.6 & 1.21  \\
GRB~040924 & 0.86 & 0.96 & 2.7  & 9.10  \\
GRB~021211 & 1.01 & 1.13 & 2.4  & 11.97 \\
GRB~050408 & 1.24 & 1.38 & 15.3 & 9.51  \\
GRB~020813 & 1.25 & 1.40 & 89.3 & 33.51 \\
GRB~060124 & 2.30 & 2.49 & 18.6 & 43.44 \\
GRB~021004 & 2.32 & 2.51 & 53.2 & 9.28  \\
GRB~030429 & 2.65 & 2.83 & 10.3 & 11.24 \\
GRB~020124 & 3.20 & 3.32 & 46.4 & 53.52 \\
GRB~060526 & 3.22 & 3.34 & 6.7  & 21.21 \\
GRB~030323 & 3.37 & 3.47 & 27.8 & 11.92 \\
\enddata
\tablecomments{T90 is defined as the time during which 5\% to 95\% of the total observed counts have been detected in the energy range 30--400~keV. The luminosity is expressed in units of $10^{51}\:\mathrm{erg}\:\mathrm{s}^{-1}$. Luminosity of GRB~050709 is not given because this burst is much shorter than the other bursts.}
\end{deluxetable}

\begin{deluxetable}{cccc}
\tabletypesize{\scriptsize}
\tablecaption{Different choices of the energy bands studied in HETE-2 GRB analysis, quoted as \textit{energy scenarios} in the text. Values of $<\Delta E>$, averages for all GRBs are given in the third column.\label{tab:ene}}
\tablewidth{0pt}
\tablehead{
\colhead{Scenario} & \colhead{Energy band 1} & \colhead{Energy band 2} & \colhead{Mean $<\Delta E>$}
}
\startdata
\#1      & 20--35 keV         & 60--350 keV     &  117.6 keV \\    
\#2      & 8--30 keV          & 60--350 keV     &  127.2 keV \\    
\#3      & 8--20 keV          & 60--350 keV     &  130.2 keV \\    
\#4      & 8--20 keV          & 30--350 keV     &   85.0 keV \\    
\#5      & 8--30 keV          & 30--350 keV     &   82.0 keV \\    
\#6      & 8--20 keV          & 40--350 keV     &  102.8 keV \\    
\#7      & 8--30 keV          & 40--350 keV     &   99.8 keV \\    
\#8      & 8--40 keV          & 40--350 keV     &   97.9 keV \\    
\#9      & 20--35 keV         & 40--350 keV     &   90.1 keV \\    
\#10     & 8--20 keV          & 50--350 keV     &  116.9 keV \\    
\#11     & 8--30 keV          & 50--350 keV     &  113.9 keV \\    
\#12     & 8--40 keV          & 50--350 keV     &  112.0 keV \\    
\#13     & 8--50 keV          & 50--350 keV     &  110.4 keV \\    
\#14     & 20--35 keV         & 50--350 keV     &  104.2 keV \\    
\enddata
\end{deluxetable}

\begin{deluxetable}{cccc}
\tabletypesize{\scriptsize}
\tablecaption{Number of pairs found for all GRBs and each energy scenario.\label{table:nbpairs}}
\tablewidth{0pt}
\tablehead{
\colhead{Scenario} & \colhead{Before cuts} & \colhead{After cuts} & \colhead{Efficiency}
}
\startdata
\#1  & 165 & 118 & 72\% \\
\#2  & 118 & 88  & 75\% \\
\#3  & 139 & 97  & 70\% \\
\#4  & 132 & 93  & 70\% \\
\#5  & 109 & 77  & 71\% \\
\#6  & 141 & 91  & 65\% \\
\#7  & 114 & 73  & 64\% \\
\#8  & 109 & 65  & 60\% \\
\#9  & 159 & 102 & 64\% \\
\#10 & 145 & 90  & 62\% \\
\#11 & 122 & 79  & 65\% \\
\#12 & 112 & 72  & 64\% \\
\#13 & 103 & 61  & 59\% \\
\#14 & 159 & 111 & 70\% \\
\hline
\multicolumn{3}{r}{Mean efficiency : } & 67\% \\
\enddata
\tablecomments{Number of pairs is obtained for both maxima and minima.}
\end{deluxetable}

\begin{deluxetable}{ccccccccccc}
\tabletypesize{\scriptsize}
\tablecaption{Results of the fit $< \Delta t > = \mathrm{a}\,K_l(z) + \mathrm{b}\,(1+z)$ for all scenarios.\label{table:resab}}
\tablewidth{0pt}
\tablehead{
         & \multicolumn{2}{c}{Minima} & \multicolumn{2}{c}{Maxima} & \multicolumn{2}{c}{All} \\
Scenario & $a$ & $b$ & $a$ & $b$ & $a$ & $b$
}
\startdata
\#1       &  0.0027$\pm$0.0116  &      0.0001$\pm$0.0070 &      -0.0780$\pm$0.0426  &       0.0613$\pm$0.0279   &    -0.0422$\pm$0.0347     &     0.0359$\pm$0.0224 \\
\#2       &  0.0487$\pm$0.0288  &     -0.0317$\pm$0.0184 &       0.0525$\pm$0.0503  &      -0.0497$\pm$0.0327   &     0.0400$\pm$0.0373     &    -0.0371$\pm$0.0240 \\
\#3       &  0.0481$\pm$0.0191  &     -0.0318$\pm$0.0116 &      -0.0034$\pm$0.0095  &       0.0048$\pm$0.0065   &     0.0169$\pm$0.0086     &    -0.0089$\pm$0.0059 \\
\#4       &  0.1144$\pm$0.0433  &     -0.0917$\pm$0.0272 &       0.0807$\pm$0.0378  &      -0.0645$\pm$0.0246   &     0.0631$\pm$0.0324     &    -0.0493$\pm$0.0218 \\
\#5       &  0.0447$\pm$0.0296  &     -0.0245$\pm$0.0182 &       0.0274$\pm$0.0530  &      -0.0330$\pm$0.0312   &     0.0409$\pm$0.0477     &    -0.0345$\pm$0.0299 \\
\#6       &  0.0812$\pm$0.0130  &     -0.0527$\pm$0.0076 &      -0.0102$\pm$0.0081  &       0.0101$\pm$0.0051   &     0.0211$\pm$0.0142     &    -0.0105$\pm$0.0101 \\
\#7       &  0.0164$\pm$0.0527  &      0.0146$\pm$0.0308 &       0.0225$\pm$0.0169  &      -0.0173$\pm$0.0117   &     0.0009$\pm$0.0383     &     0.0105$\pm$0.0243 \\
\#8       &  0.1056$\pm$0.0567  &     -0.0449$\pm$0.0308 &      -0.0283$\pm$0.0160  &       0.0122$\pm$0.0095   &     0.0159$\pm$0.0142     &    -0.0139$\pm$0.0100 \\
\#9       &  0.0651$\pm$0.0327  &     -0.0524$\pm$0.0187 &       0.0286$\pm$0.0406  &      -0.0272$\pm$0.0246   &    -0.0044$\pm$0.0149     &     0.0058$\pm$0.0097 \\
\#10      &  0.0756$\pm$0.0330  &     -0.0338$\pm$0.0188 &      -0.0011$\pm$0.0089  &       0.0080$\pm$0.0059   &     0.0139$\pm$0.0296     &     0.0009$\pm$0.0195 \\
\#11      &  0.0493$\pm$0.0274  &     -0.0263$\pm$0.0179 &       0.0341$\pm$0.0251  &      -0.0242$\pm$0.0163   &     0.0584$\pm$0.0322     &    -0.0389$\pm$0.0220 \\
\#12      &  0.0353$\pm$0.0305  &     -0.0246$\pm$0.0198 &      -0.0373$\pm$0.0308  &       0.0379$\pm$0.0183   &     0.0209$\pm$0.0207     &    -0.0136$\pm$0.0142 \\
\#13      &  0.1001$\pm$0.0493  &     -0.0417$\pm$0.0271 &       0.0313$\pm$0.0126  &      -0.0206$\pm$0.0078   &    -0.0366$\pm$0.0432     &     0.0348$\pm$0.0280 \\
\#14      &  0.0604$\pm$0.0525  &     -0.0117$\pm$0.0301 &      -0.0444$\pm$0.0385  &       0.0347$\pm$0.0257   &    -0.0358$\pm$0.0286     &     0.0310$\pm$0.0188 \\
\enddata
\tablecomments{Errors are obtained from the fit and normalised to $\chi^2/ndf \sim1$.}
\end{deluxetable}

\begin{deluxetable}{cccc}
\tabletypesize{\scriptsize}
\tablecaption{Obtained 95\%~CL lower limit on $\mathrm{E}_\mathrm{QG}$ (GeV).\label{table:reseqg}}
\tablewidth{0pt}
\tablehead{
\colhead{Scenario} & \colhead{Minima} & \colhead{Maxima} & \colhead{All}
}
\startdata
\#1	   &  $5.9\times10^{14}$   &  $5.8\times10^{14}$	& $5.7\times10^{14}$ \\
\#2	   &  $5.8\times10^{14}$   &  $7.8\times10^{14}$	& $9.0\times10^{14}$ \\
\#3	   &  $7.5\times10^{14}$   &  $3.2\times10^{15}$	& $2.0\times10^{15}$ \\
\#4	   &  $3.6\times10^{14}$   &  $5.8\times10^{14}$	& $6.3\times10^{14}$ \\
\#5	   &  $3.8\times10^{14}$   &  $4.6\times10^{14}$	& $4.6\times10^{14}$ \\
\#6	   &  $4.7\times10^{14}$   &  $1.8\times10^{15}$	& $1.3\times10^{15}$ \\
\#7	   &  $2.3\times10^{14}$   &  $1.7\times10^{15}$	& $4.5\times10^{14}$ \\
\#8	   &  $2.1\times10^{14}$   &  $1.2\times10^{15}$  & $1.8\times10^{15}$ \\
\#9	   &  $4.0\times10^{14}$   &  $5.8\times10^{14}$	& $1.0\times10^{15}$ \\
\#10   &  $3.6\times10^{14}$   &  $1.9\times10^{15}$	& $7.3\times10^{14}$ \\
\#11   &  $5.3\times10^{14}$   &  $8.6\times10^{14}$	& $7.9\times10^{14}$ \\
\#12	 &  $5.8\times10^{14}$   &  $5.0\times10^{14}$	& $1.2\times10^{15}$ \\
\#13	 &  $2.4\times10^{14}$   &  $1.2\times10^{15}$	& $5.2\times10^{14}$ \\
\#14	 & 	$2.3\times10^{14}$   &  $7.7\times10^{14}$	& $6.8\times10^{14}$ \\
\enddata
\end{deluxetable}

\clearpage

\end{document}